*Chapter 2*

# BI CONTAINING MULTIFERROIC PEROVSKITE OXIDE THIN FILMS

*Rainer Schmidt [1,\*], Eric Langenberg [2,3], Jofre Ventura [2]
and Manuel Varela [2]*

[1]Universidad Complutense de Madrid, GFMC, Dpto. Física Aplicada III, Facultad de Ciencias Físicas, Madrid, Spain

[2]Universitat de Barcelona, Dpto. Física Aplicada i Óptica, Diagonal Sud, Facultats de Física i Química, Barcelona, Spain

[3]Universidad de Zaragoza, Instituto de Nanociencia de Aragón, Mariano Esquillor, Zaragoza, Spain

## ABSTRACT

In this work multiferroic thin films of Bi containing perovskite oxides are discussed, where the driving force for ferroelectricity are the $Bi^{3+}$ lone-pair electrons. First, a brief introduction of Bi containing multiferroic perovskite oxides will be presented to describe the mechanisms for establishing magnetic and ferroelectric order in these compounds and some recent developments in this field of research are reported. The second section addresses experimental aspects of epitaxial thin film growth of $BiMnO_3$, $(Bi_{0.9}La_{0.1})_2NiMnO_6$ (BLNMO) and $BiFeO_3$ thin films by pulsed laser deposition. The third section is dedicated to the physical properties of such films in terms of structural characterization and the magnetic and ferroelectric properties and their correlations in form of magnetoelectric coupling (MEC).

[\*] Corresponding author. Email: rainerxschmidt@googlemail.com.

# 1. INTRODUCTION

## 1.1. Pathways to the Coexistence of Magnetism and Ferroelectricity in Perovskites

Multiferroics are materials where at least two ferroic orders coexist in the same phase. The definition of the term ferroic aims to characterise materials in which domains are formed with a corresponding net macroscopic magnitude that can be hysteretically switched by applying an external field, i.e. ferroelectric, ferromagnetic, ferroelastic and ferrotoroidic orders (primary ferroics) [1, 2, 3]. Nonetheless, antiferroic orders consisting in antiferromagnetism, ferrimagnetism and antiferroelectricity are widely accepted to be also included in the extended definition of the term multiferroic. Among all kinds of multiferroics, those showing both ferroelectric and magnetic order draw special attention in terms of their magnetoelectric properties, which is the focus of this work. The particular interest in such multiferroics is driven by industrial aspirations to design charge storage devices, where information could be written electrically onto a data bit and retrieved magnetically. For the sake of simplicity the term "multiferroics" is used here to refer to ferroelectric and (anti-) ferromagnetic multiferroics.

For multiferroic charge storage devices it is required that the magnetization (polarization) can be modified by applying an electric (magnetic) field and vice versa, which implies that ferroelectric and magnetic orders need to be strongly coupled to each other (Figure 1). The so-called magnetoelectric coupling (MEC) effect has therefore great technological potential for novel magnetoelectric applications and spintronics [4-8]. Yet it is worth mentioning that the sole coexistence of the two ferroic orders is not a sufficient condition for MEC to occur, which is a more restrictive material property.

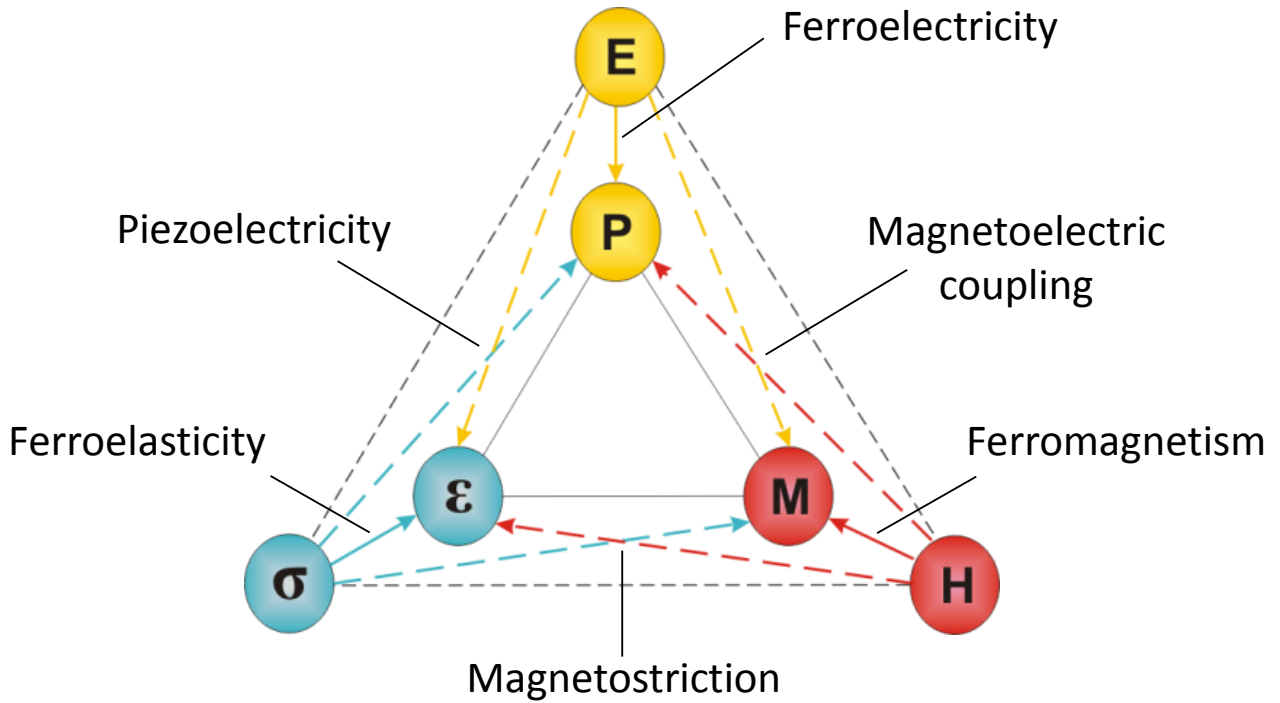

Figure 1. Summary of all the functionalities that multiferroic materials may offer. $E$ and $P$ stand for electric field and electric polarization, $H$ and $M$ stand for magnetic field and magnetization, and $\sigma$ and $\varepsilon$ stand for stress and deformation, respectively. Adapted from Reference [3].

Despite the large number of ferroelectric or magnetic materials existing in nature, the combination of ferroelectric and ferromagnetic orders in one intrinsic material seems to be quite elusive [9-11]. Whereas magnetic materials can be either insulating or conductive, ferroelectric materials can solely be insulating, otherwise an applied electric field would induce an electric current rather than switching ferroelectric domains. Second, from crystal symmetry considerations (see Figure 2) ferroelectricity breaks the spatial inversion symmetry, which requires non-centrosymmetric crystal structures for ferroelectric order to occur. Conversely, magnetism breaks time-reversal symmetry [2]. Combination of the two symmetry restrictions leaves only 13 point groups (among 233) that would enable ferroelectricity and magnetism to coexist [9-11].

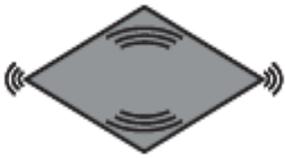

Figure 2. Ferroic orders under inversion symmetry of time and space [2].

Since most ferroelectrics and a large number of magnetic materials are transition metal oxides with the $ABO_3$ perovskite structure (e.g. $BaTiO_3$ and $(La, Sr)MnO_3$), much of the attention was reasonably drawn to this crystal structure type for the search for multiferroic materials. Both, in ferroelectric and magnetic perovskites it is usually the B-cation which drives the ferroelectric displacement or the magnetic order, respectively. In the case of ferroelectrics, the B-cation displaces from the centre of the surrounding oxygen octaheadra, breaking the centrosymmetry and creating an electric dipole. Yet this condition requires the B-cation to have a $d^0$ electron configuration, e.g. empty $d$ orbitals like in $Ti^{4+}$, which minimizes the Coulombian electrostatic repulsion of the surrounding oxygen anions [9]. Magnetism in transition metal perovskites arises from superexchange interactions (or double exchange for conductive magnetic oxides) between the B-cations mediated by the adjacent oxygen anions [9, 12]. This condition requires the B-cation to possess a magnetic moment, which is only possible when the $d$ orbitals of the B-cation are partially occupied. Thus, a mutual exclusion between ferroelectricity and magnetism in perovskites exists to some extent [9].

To date, most attempts to design multiferroic materials have been based on looking for new sources for ferroelectricity, while maintaining the same recipes for magnetism. Based on such considerations, multiferroics have been classified in two types as described by Khomskii [13]: In Type-I multiferroics, ferroelectricity and magnetism rely on two independent mechanisms, whereas in Type-II the ferroelectricity arises from the magnetic order, i.e. ferroelectricity exists only in a magnetically ordered state. The latter is expected to produce large magnetoelectric coupling as ferroelectric order is intrinsically related to the magnetic one, but such ferroic properties tend to order at rather low temperatures. This coupled type of ferroelectric and magnetic order often involves commensurate and incommensurate spiral spin structures, which induce ferroelectricity. This occurs in several rare-earth manganites [14-16]. Contrarily, Type-I multiferroics tend to show much higher ferroic transition temperatures at the expense of a small MEC.

### 1.1.1. Ferroelectricity in the Presence of $Bi^{3+}$ Lone-pair Electrons

The work presented here focuses on one particular subgroup of Type-I multiferroics, where ferroelectricity is induced by means of lone-pair electrons. In order to achieve the two ferroic orders in $ABO_3$ perovskite oxides it is obvious that one of the cations may be utilized for inducing ferroelectric order, while the other may induce magnetism. One possibility is the exploitation of a $d^0$ transition metal cation located on the *B*-site and a magnetic cation on the A-site, for example in $EuTiO_3$ [17]. The second possibility is the use of stereochemical active cations like $Bi^{3+}$ or $Pb^{2+}$ on the perovskite A-site inducing ferroelectricity due to lone-pair electrons, and a magnetic cation located on the B-site. The latter approach applies to all systems studied here, $BiMnO_3$, $(Bi_{0.9}La_{0.1})_2NiMnO_6$ and $BiFeO_3$. The electronic configuration of $Bi^{3+}$ (and also $Pb^{2+}$) is $[Xe]4f^{14}5d^{10}6s^26p^0$ with empty $6p$-states and the two $6s$ outer lone-pair electrons do not participate in chemical bonds.

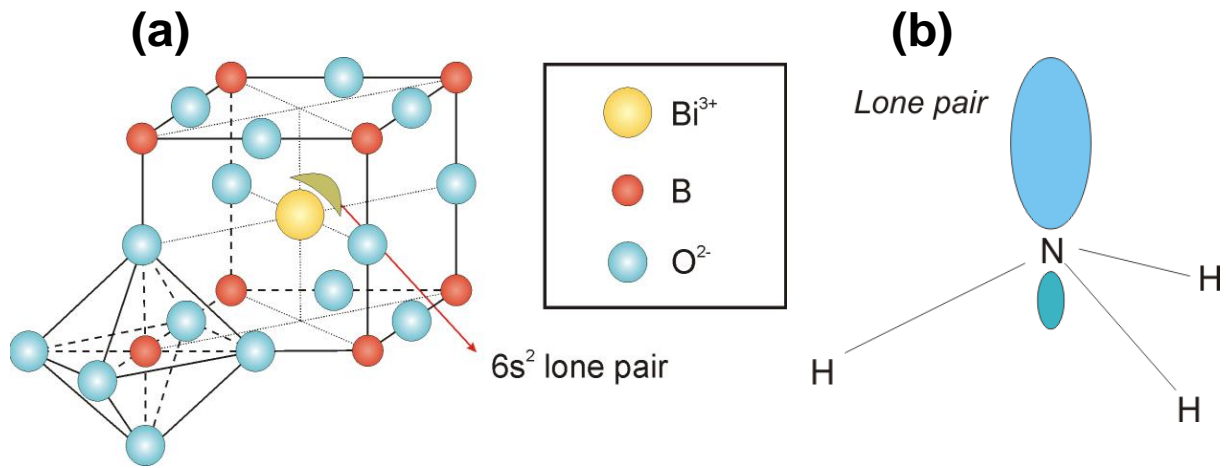

Figure 3. (a) Schematic representation of the lobe-like distribution of the $6s^2$ lone-pair electrons in Bi-based perovskite structures, $BiBO_3$, breaking the spatial inversion symmetry. (b) Lone-pair $2s^2$ in the ammonia molecule.

In the absence of interactions, the lone-pairs are nearly spherically distributed, but when surrounded by oxygen anions they shift away from the centrosymmetric position due to the Coulombian electrostatic repulsion forming a localized lobe-like distribution, very much alike the ammonia molecule (Figure 3) [18]. Thus, the lone-pairs form an electric dipole, which breaks the spatial inversion symmetry. This mechanism is the driving force for ferroelectricity in all Bi-based multiferroic perovskites.

### *1.1.2. Magnetic Ordering in Bi Containing Perovskite Oxides*

If $Bi^{3+}$ is situated on the perovskite A-site, this allows the location of a magnetic transition metal oxide cation on the B-site with a (partially) occupied outer electron shell. Since B-site cation chains are interrupted by oxygen anions, long range magnetism arises from magnetic superexchange interactions mediated by the adjacent oxygens: B - O - B.

The surrounding oxygen octahedra of the B-site splits *d*-levels into two energy states: two high-energetic $e_g$ and three low-energy $t_{2g}$ orbitals [19]. Whether the magnetic interaction is ferromagnetic or antiferromagnetic depends on the filling of the $e_g$ orbitals according to the Goodenough-Kanamori's (GK) rules (Fig.4).

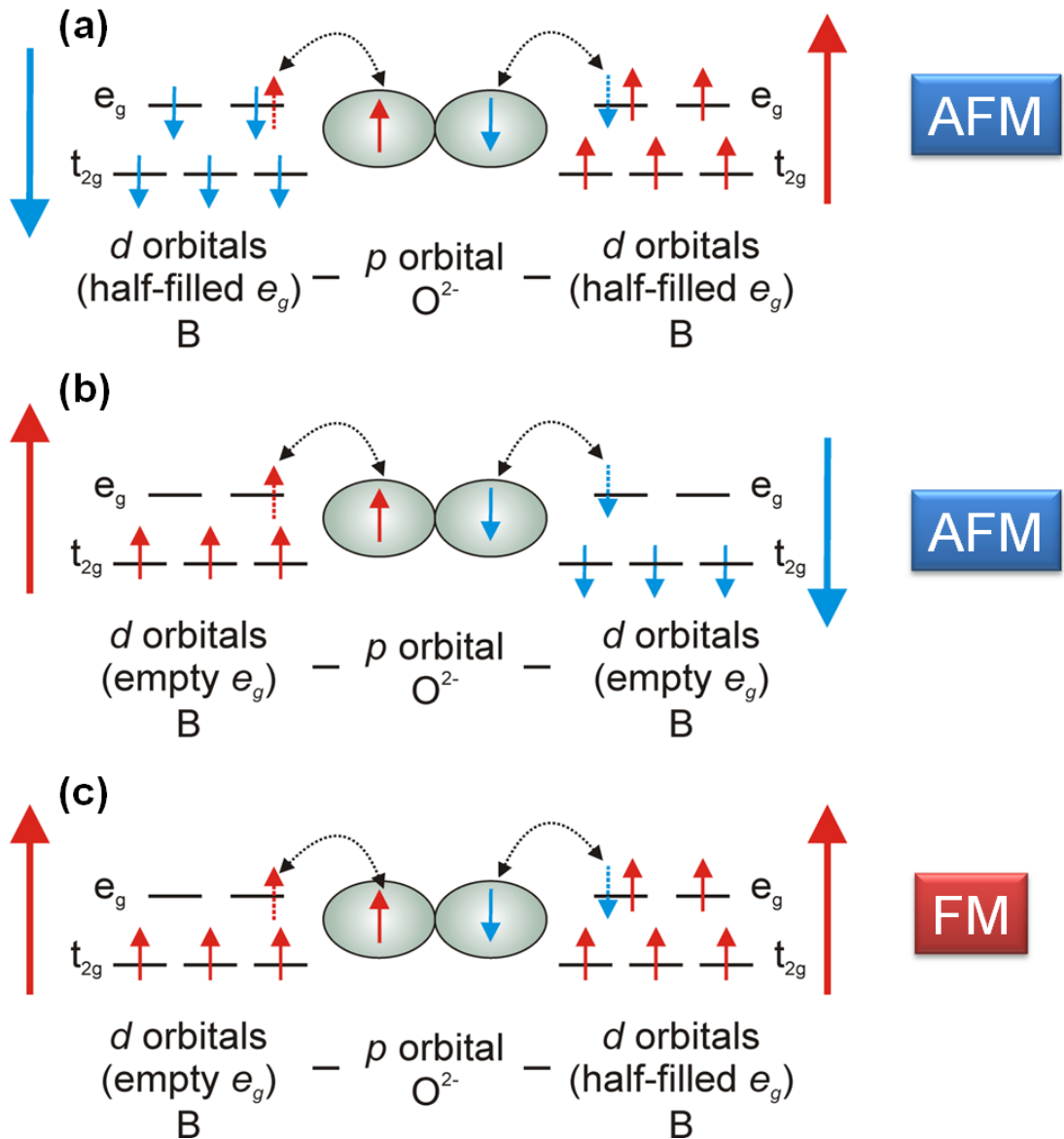

Figure 4. Schematic representation of superxchange interaction in $ABO_3$ perovskite with an ideal $180°$ B - O - B bonding-angle. Depending on the occupation of the $e_g$ orbitals, the type of magnetic interaction is antiferromagnetic (a, b) or ferromagnetic (c).

Bearing in mind *(i)* the Pauli's exclusion principle stating that two electrons in the same orbital must possess antiparallel spins, and *(ii)* the Hund's rule stating that electrons in degenerated *d* orbitals minimize their energy in a parallel spin alignment, Goodenough-Kanamori's rules can be summarized as follows:

a) Either B (empty $e_g$ orbitals) - O - B (empty $e_g$ orbitals) or B (half-filled $e_g$ orbitals) - O - B (half-filled $e_g$ orbitals) give rise to antiferromagnetic interactions (Figure 4a, b).
b) B (empty $e_g$ orbitals) - O - B (half-filled $e_g$ orbitals) give rise to ferromagnetic interactions (Figure 4c).

Note that these GK rules are deduced in the simplest scenario of an ideal perovskite where the B - O - B bond angle is 180º. Distorted perovskites, rotation of the oxygen octahedral and different B - O - B bond angles may give rise to different magnetic interactions [12].

In the case that the perovskite B-site is occupied by only one kind of cation with the same oxidation state and the same $e_g$ orbital filling, only antiferromagnetic interactions can occur and all multiferroic $BiBO_3$ with the B-site occupied by the same cation are antiferromagnetic.

In order to induce ferromagnetism, two different magnetic transition metal cations occupying the B-site in a double perovskites oxide, $Bi_2BB'O_6$, can provide ferromagnetic exchange according to the GK rules, but only if the different cations are long-range ordered in an alternating fashion. The appropriate choice of B - B' cations should therefore be aimed at such B-site cation long-range order to be likely and the following electron configuration: B (empty $e_g$ orbitals) - O - B' (half-filled $e_g$ orbitals).

Still, if the antiferromagnetic interactions dominate in a specific B-site ordered double-perovskite, long range ferrimagnetism is likely to occur due to the fact that every transition metal oxide possesses different magnetic moment (different *d* orbital filling), where the magnetic moments are not compensated. For this reason, Bi-based double perovskite oxides tend to be either ferromagnetic or ferrimagnetic, showing net magnetization. However, it is worth noting that due to the different electronic configurations of the B - B' cations, double-perovskite oxides tend to be less insulating than their counterpart 'single' perovskite oxides.

## 1.2. Bi-Based Single Perovskites, $BiBO_3$

Almost all $BiBO_3$ single perovskite multiferroics are antiferromagnetic, with the unique exception of $BiMnO_3$. Therefore, $BiMnO_3$ is one of the most prominent multiferroics together with $BiFeO_3$, where the latter is the only known single phase compound which is ferroelectric and antiferromagnetic at room temperature. These two most important Bi containing single perovskite multiferroics will be reviewed in detail here. Furthermore, several more recently discovered Bi-based single perovskite oxides will be mentioned briefly.

### *1.2.1. $BiMnO_3$*

$BiMnO_3$ is commonly synthesized in bulk form at ≈ 600 - 700 ºC under high pressure of ≈ 3 - 6 GPa to stabilize the phase [20-24]. Single-phase $BiMnO_3$ can also be obtained in thin film form by pulsed laser deposition, where the high-pressure requirement for phase stabilization is replaced by the epitaxial strain imposed by the substrate [25-28]. Still, the high volatility of Bi together with the multiphase tendency of the Bi - Mn - O ternary system entails difficulty to obtain single-phase stabilization even in thin films [29-32].

$BiMnO_3$ is expected to crystallize in an orthorombic structure similar to $LaMnO_3$ since $La^{3+}$ and $Bi^{3+}$ have similar cationic radii [33]. However, this is not the case at low temperatures, where $BiMnO_3$ crystallizes in a highly distorted non-centrosymmetric monoclinic *C2* structure [20] due to the stereochemical activity of $Bi^{3+}$ lone-pairs [18]. On increasing temperature $BiMnO_3$ undergoes two structural phase transitions at ≈ 450 K and ≈ 770 K [22]. Between ≈ 450 - 770 K the structure can still be indexed as non-centrosymmetric monoclinic *C2*, whereas above ≈ 770 K the structure is centrosymmetric *Pbnm* which no longer allows spontaneous polarization (Figure 5).

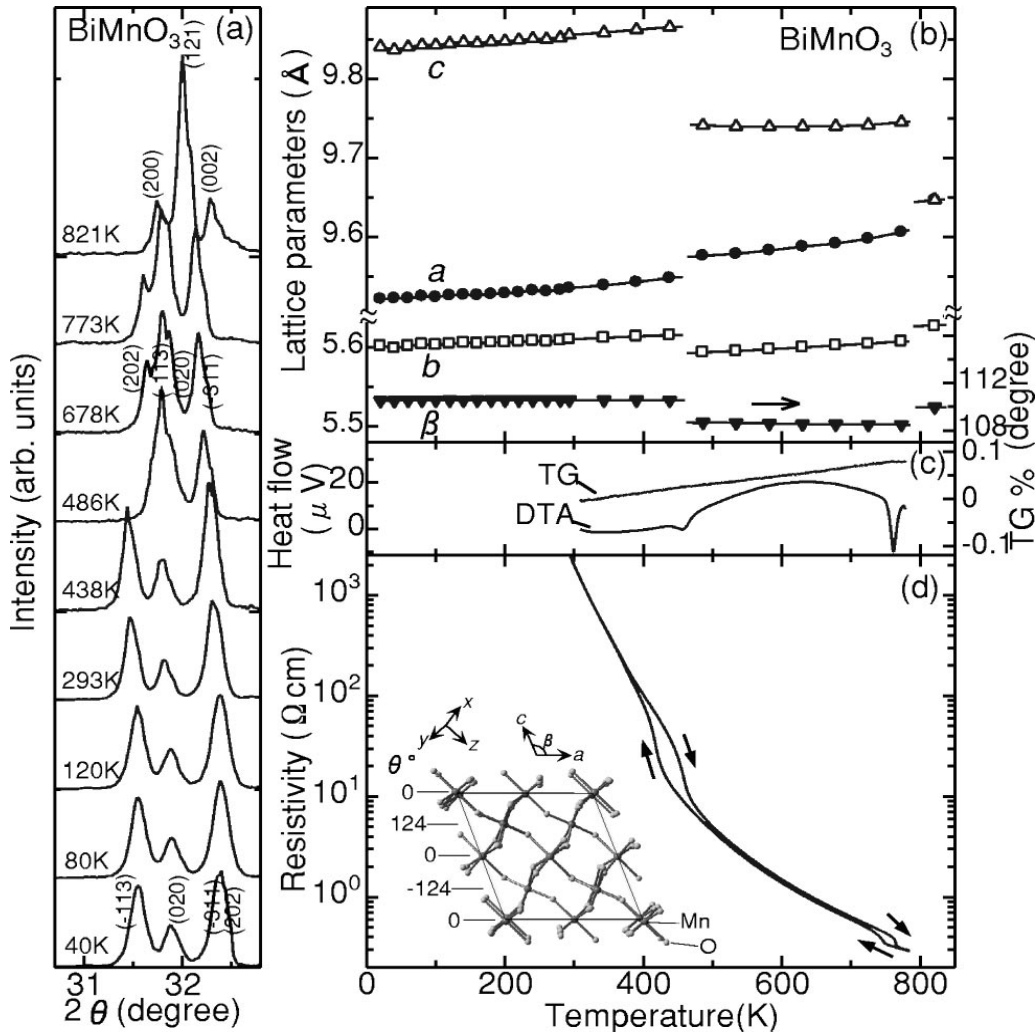

Figure 5. (a) Powder X-ray diffraction of $BiMnO_3$ at different temperatures. (b) Temperature dependence of lattice parameters. Above ≈ 770 K the structure can be indexed as centro-symmetric Pbnm. (c) Thermogravimetry and differential thermal analysis showing two phase transitions. (d) Temperature dependence of resistivity signalling the two phase transitions. Reproduced from Reference [22], with permission from the American Physical Society (APS).

Thus, $BiMnO_3$ is considered to be ferroelectric up to ≈ 770 K well above room temperature, although neither conclusive ferroelectric hysteresis loops nor ferroelectric domain switching current measurements have been reported due to high leakage. In fact, some revised analysis of the crystal structure [34, 35] and first principles calculations [36] have questioned the non-centrosymmetric structure of $BiMnO_3$. However, the four-resistive states reported for La-doped $BiMnO_3$ multiferroic tunnel junctions [4] can only be understood in the framework of a ferromagnetic and ferroelectric material, which strongly indicates ferroelectric behaviour at least in thin film form. In addition, nonlinear optical measurements on $BiMnO_3$ thin films have revealed changes in the polar symmetry of the second harmonic generation signal by applying electric fields, consistent with changes in the ferroelectric domain structure [37]. Nevertheless, a clear-cut evidence for switching ferroelectric domains by the application of electric fields is still missing, despite strong indications for ferroelectricity.

The ferromagnetic order in $BiMnO_3$ sets in below ≈ 105 K [22]. Despite the similar orbital ordering compared to antiferromagnetic $LaMnO_3$, two out of three Mn - O - Mn orbital configurations in $BiMnO_3$ favour ferromagnetic interactions, which results in an overall long-range ferromagnetism, overcoming the antiferromagnetic interactions [22]. It should be noted that the Mn - O - Mn bond angles in the highly distorted monoclinic structure of $BiMnO_3$ are significantly smaller than the ideal 180º (≈ 140 - 160º) which favours ferromagnetism in this case [22]. In bulk the saturated magnetization is found to be ≈ 3.6 Bohr magneton ($\mu_B$) per formula unit (f.u.) [22], whereas lower values were reported in thin films (≈ 2.2$\mu_B$/f.u.) [27].

### *1.2.2. BiFeO₃*

No other multiferroic material has been more extensively studied than $BiFeO_3$, because its multiferroic properties are robustly maintained well above room temperature. The crystal structure of $BiFeO_3$ is rhombohedral *R3c* [38, 39], which is a polar group and thus permits the establishment of ferroelectric order.

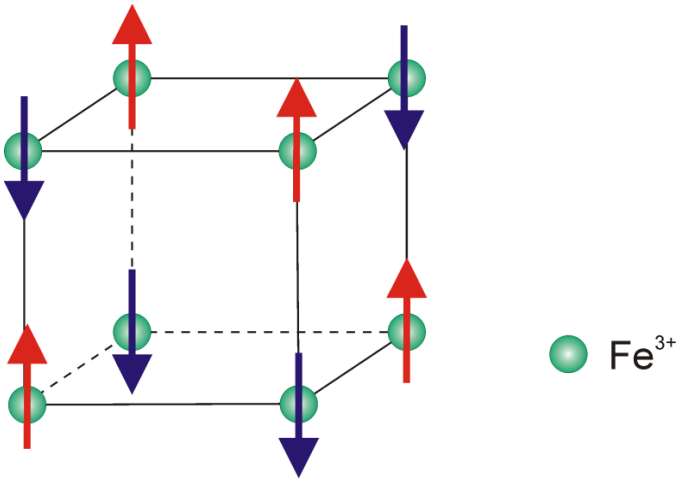

Figure 6. The magnetic structure of $BiFeO_3$: G-type antiferromagnetism.

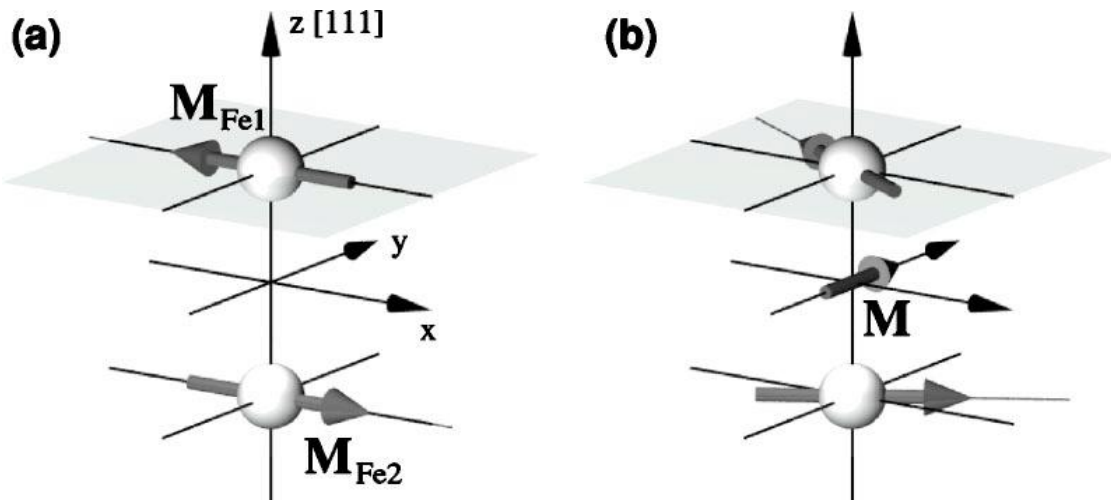

Figure 7. Weak ferromagnetism in $BiFeO_3$. Reproduced from Reference [44], with permission from the American Physical Society (APS).

The ferroelectric transition temperature was proven to be rather high, $T_{FE}$ = 1103 K, above which a centrosymmetric paraelectric phase with space group $R\bar{3}c$ can be indexed [40]. In the ferroelectric regime the polar axis is found along the [111] direction and a large saturated polarization, $\approx$ 80 - 100 $\mu$C/cm$^2$, is reported as expected for a high-Curie temperature ferroelectric [41]. BiFeO$_3$ exhibits antiferromagnetism with G-type arrangement (Figure 6) and a Néel temperature of $\approx$ 620 K [42]. Additionally, superimposed on the antiferromagnetic order, an incommensurate cycloidal spiral arrangement of the spins with a long period of 620 Å was reported [43]. Moreover, the symmetry of the BiFeO$_3$ structure allows small canting of the spins to occur (Figure 7), which would give rise to a net magnetization in terms of weak ferrimagnetism of the Dzyaloshinski-Moriya type [44]. This weak ferrimagnetism requires the spiral spin structure to be suppressed, which can be attained in thin films due to the epitaxial constraints or enhanced anisotropy [44]. The expected magnetization would be $\approx$ 0.1 $\mu_B$/f.u., which was confirmed experimentally. Despite early claims of higher magnetisation [41], the true value is now quite well established to be close to the theoretical 0.1 $\mu_B$/f.u. [45].

Most importantly, by combining photoelectron emission microscopy and piezoforce microscopy it was shown that the antiferromagnetic and ferroelectric domains in BiFeO$_3$ are strongly coupled [46], which was confirmed by neutron scattering experiments on single crystals [47]. The magnetoelectric coupling was argued to be diminished in thin films because of the absence of the spiral cycloid of the spin arrangement [47].

### *1.2.3. Other Bi-Based Single Perovskites*

Other Bi-based single perovskites have been proposed and proved to be multiferroic, but all of them are antiferromagnetic with only weak ferrimagnetism at most. First principles calculations predicted BiCrO$_3$ to be antiferroelectric and antiferromagnetic [48]. Experimental results show weak ferromagnetic ordering ($\approx$ 0.05 $\mu_B$/f.u. saturated magnetization) below $\approx$ 120 K in bulk [49] and thin film samples [50], but contradictory claims were made about whether BiCrO$_3$ is antiferroelectric [51] or ferroelectric [50]. Another potential multiferroic is BiCoO$_3$, which was predicted to crystallize in a tetragonal structure with large tetragonality *c/a* giving rise to large polarization ($\approx$ 150 $\mu$C/cm$^2$) [52]. The structure was confirmed by neutron diffraction [53], though the ferroelectric character has not been experimentally detected yet. Furthermore, BiNiO$_3$ has been considered, which is antiferromagnetic with a weak spin canted ferrimagnetic moment below 300 K as a consequence of the small Ni - O - Ni bond angles [54]. Due to the preferred Ni$^{2+}$ valence state, BiNiO$_3$ displays an unusual charge disproportionation at the A-site of Bi$^{3+}$ - Bi$^{5+}$. This is quite rare for Bi cations and affects the magnetic symmetry in BiNiO$_3$ [54]. Still, there is no experimental evidence of its ferroelectric character and indeed structural analysis reveals BiNiO$_3$ to be centrosymmetric [55].

## 1.3. Bi-Based Double-perovskites, Bi$_2$BB'O$_6$

Bi-based double-perovskites Bi$_2$BB'O$_6$ have been much less studied than their counterpart single perovskites, but recently more intense research has resulted in promising findings such as a small net magnetization at room temperature in Bi$_2$FeMnO$_6$ [56]. As mentioned before, the Goodenough-Kanamori's rules predict these oxides to allow engineering of ferromagnetic pathways. In an ideal ferromagnetic double-perovskite the B - B' cation species may possess either $d^3$ - $d^5$ or $d^3$ - $d^8$ configuration, and an ideal B - O - B' bonding angle of 180º [12]. However, due to the mixed valence states that most transition metal oxides can adapt, the desired electronic configuration of B - B' cations has been found difficult to be obtained. Moreover, long-range ferromagnetism requires alternating B - B' cation ordering along the fundamental perovskite directions [104], [010] and [001] as illustrated in Figure 8. Among all double-perovskites, Bi$_2$NiMnO$_6$ has been the most intensively studied due to the fact that B-site Ni$^{2+}$ and Mn$^{4+}$ cations have been proven to be long-range ordered [57, 58], thus allowing long-range ferromagnetism to occur. In the following, Bi$_2$NiMnO$_6$ will be discussed in detail and several alternative Bi-based double perovskites will be mentioned briefly.

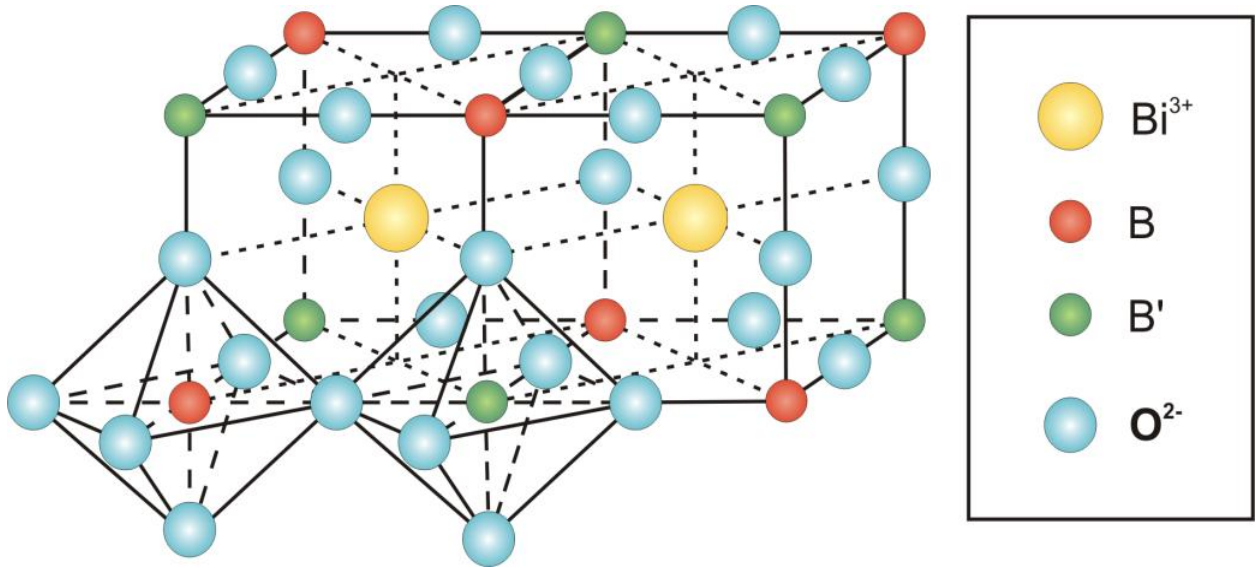

Figure 8. Scheme of the double-perovskite structure with long range B-site cation ordering.

### 1.3.1. $Bi_2NiMnO_6$

Equivalent to $BiMnO_3$, $Bi_2NiMnO_6$ is commonly synthesized in bulk form at ≈ 800ºC under high pressure of ≈ 6 GPa [59]. Yet epitaxial strain was also proven to be equally efficient, replacing the high synthesis pressures to produce single phase samples by pulsed laser deposition [60-62]. However, much alike $BiMnO_3$, the high Bi volatility and the multiphase tendency of these materials lead to narrow windows in the thin film deposition conditions for single phase stabilization [62]. On the other hand, partial replacement of $Bi^{3+}$ cations by $La^{3+}$ cations on the A-site has been proven to facilitate single-phase stabilisation in bismuth manganite compounds [30, 62-64] due to the slightly smaller ionic radius of $La^{3+}$ with regard to $Bi^{3+}$ (0.130 nm and 0.131 nm, respectively) [33]. It gives rise to a slightly reduced unit cell volume and exerts chemical pressure to prevent $Bi^{3+}$ cation desorption during film growth.

$Bi_2NiMnO_6$ crystallizes in the monoclinic $C2$ structure, which is non-centrosymmetric allowing spontaneous polarization. The ferroelectric phase transition occurs at ≈ 485 K, above which the compound crystallizes in a centrosymmetric monoclinic $P2_1/n$ structure [59] and a dielectric anomaly can be observed across the transition (Figure 9a) [59].

The solid solution $(Bi_{1-x}La_x)_2NiMnO_6$ remains non-centrosymmetric at sufficiently low temperature for $x <$ 0.2 [65] allowing ferroelectric order to be established. Conclusive measurements to detect ferroelectric domain switching current on 10% La-doped $(Bi_{0.9}La_{0.1})_2NiMnO_6$ thin films have confirmed the ferroelectric character of this compound [66], although the ferroelectric Curie temperature (≈ 450 K) was inferred to be slightly inferior than that in bulk (≈ 485 K).

As stated above, ferromagnetism in $Bi_2NiMnO_6$ arises from the long-range order of $Ni^{2+}$ - O - $Mn^{4+}$ cations. The preferred $Ni^{2+}$ valence state forces the Mn cation to adapt 4+ valency, which not only favours the appropriate electronic configuration of the B cations for ferromagnetic interactions, but also the long range B-site cation order. $Bi_2NiMnO_6$ shows a large saturated magnetization (≈ 4 $\mu_B$/f.u.) and orders ferromagnetically below 140 K in bulk (Figure 9b) [59], whereas a lower Curie temperature (≈ 100 K) was reported for thin films [58, 60].

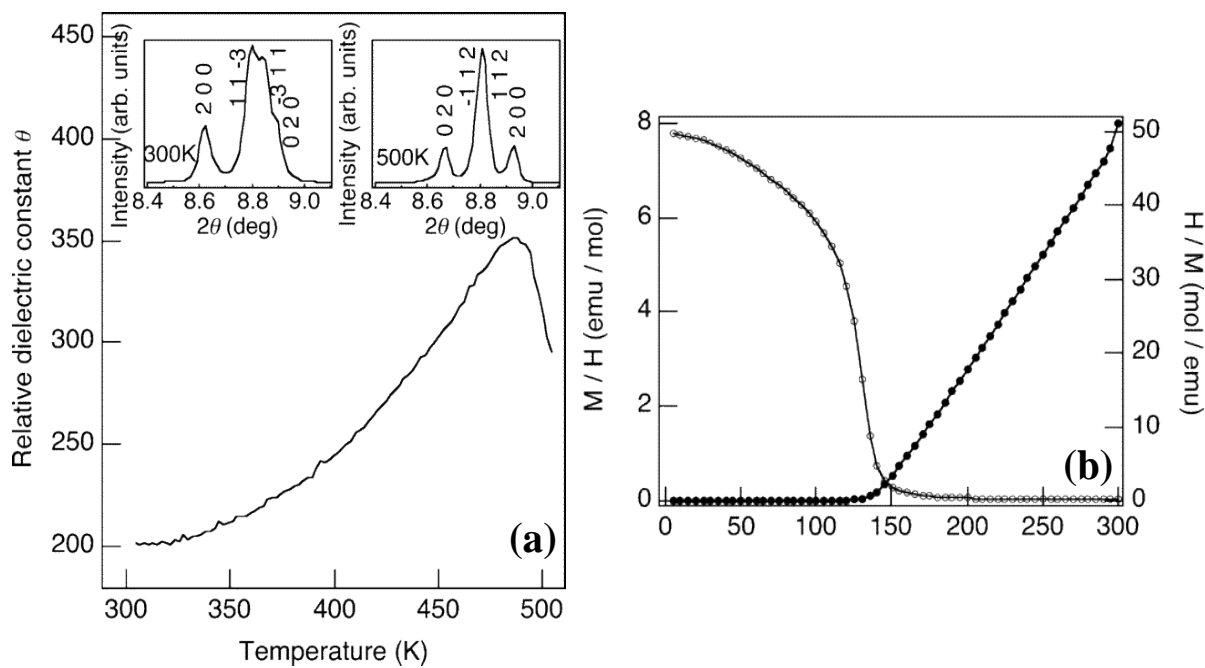

Figure 9. Temperature dependence of (a) the dielectric permittivity and (b) the magnetization of bulk $Bi_2NiMnO_6$. Reproduced from Reference [59], with permission from the American Chemical Society (ACS).

### 1.3.2. Other Bi-Based Double-perovskites

Other Bi-based double perovskites have been studied and proved to be multiferroic, but none of them show clear long-range B-site cation order, long-range ferromagnetism or ferrimagnetism.

As a first example, $Bi_2FeMnO_6$ is predicted to crystallize in monoclinic *C2* structure [67], but bulk samples are usually indexed as orthorhombic [68]. Both Fe and Mn are found to be mainly in a 3+ valence state, which hinders the B-site cation ordering [56, 67, 68]. A hypothetical B-site ordering would entail either ferromagnetic or more likely antiferromagnetic exchange with a net ferrimagnetic magnetization of ≈ 1 $\mu_B$/f.u. (the magnetic moments of $Fe^{3+}$ and $Mn^{4+}$ are 5 $\mu_B$ and 4 $\mu_B$, respectively). Yet the recorded magnetization values are much lower, thus confirming the lack of B-site order in this compound [56, 67, 68]. On the other hand, $Bi_2FeMnO_6$ was proven to be ferroelectric with a saturated polarization of ≈ 30 $\mu C/cm^2$ at 150 K [69].

Another example is $Bi_2FeCrO_6$. Ab-initio calculations predicted $Bi_2FeCrO_6$ to show a large saturated polarization (≈ 80 $\mu C/cm^2$) and a saturated magnetization of 2 $\mu_B$/f.u. for a $Cr^{3+}$ and $Fe^{3+}$ B-site cation order [70]. Although $Cr^{3+}$ and $Fe^{3+}$ exhibit $d^3$ and $d^5$ orbital filling and ferromagnetic superexchange interaction would be expected, the magnetic $Fe^{3+}$ - O - $Cr^{3+}$ interaction was predicted to couple antiferromagnetically [70]. Still, a net magnetization of 2 $\mu_B$/f.u. would be expected in case of long-range cation order due to the uncompensated magnetic moments (5$\mu_B$ and 3$\mu_B$ for $Fe^{3+}$ and $Cr^{3+}$, respectively). However, long-range B-site order in $Bi_2FeCrO_6$ is not likely to occur due to the same charge of $Fe^{3+}$ and $Cr^{3+}$ and the similar ionic radii (0.645 Å and 0.615 Å, respectively) [33]. Indeed $Bi_2FeCrO_6$ is found to exhibit low magnetization (< 0.2 $\mu_B$/f.u.) in both bulk samples [71] and thin films [72]. Nonetheless, some recent experimental results show a certain thickness dependence of the B-site order, pointing out that partial ordering of $Fe^{3+}$ and $Cr^{3+}$ can be improved by decreasing the thickness of $Bi_2FeCrO_6$ thin films [73]. The predicted ferroelectric properties were confirmed experimentally with a large polarization of ≈ 60 $\mu C/cm^2$ along the [001] direction at ≈ 77 K, which would entail ≈ 100 $\mu C/cm^2$ along the polar axis [111], [72].

To date all known Bi containing multiferroic double perovskite oxides consist of $3d$ transition metals occupying the B-site, like Ni, Mn and Fe. Recently it was predicted though that combining $3d$ and $5d$ elements in Bi-based double perovskites allows robust room temperature ferrimagnetism to be attained [74], which is still a big challenge in multiferroics for the time being. In particular, $Bi_2NiReO_6$ and $Bi_2MnReO_6$ are expected to order ferrimagnetically below 360 K and 330 K, respectively. Although they would not be ferroelectric in the ground state, the ferroelectric phase is likely to be stabilized in thin films by small percentages of strain [74]. Still, these materials have not been synthesized so far, hence their predicted multiferroic properties remain to be confirmed experimentally.

## 2. EXPERIMENTAL

Functional properties in epitaxial thin films strongly depend on the film quality and the lattice strain. The latter arises due to the lattice mismatch between the substrate and the bulk crystal of the epitaxial phase under investigation. Therefore, the deposition technique used and the choice of an appropriate substrate are of paramount importance for the production of high quality Bi containing multiferroic perovskite thin films, as will be demonstrated on the examples of $BiMnO_3$, $(Bi_{0.9}La_{0.1})_2NiMnO_6$ (BLNMO) and $BiFeO_3$.

### 2.1. Pulsed Laser Deposition of Bi-Based Perovskite Multiferroics

Among other physical vapor deposition techniques, pulsed laser deposition (PLD) is long known as the tool of choice for the growth of complex-oxide materials [75]. The technique is conceptually simple, even though involving rather complex physical processes: a pulsed laser beam leads to the ejection of particles from a solid target under conditions far away from thermal equilibrium, where the particles range from atoms to electrons and clusters of particles. Such particles form a spatially restricted plasma known as plume, which propagates in the evacuated deposition chamber towards the substrate and condensates on it. Figure 10 illustrates the top view of the PLD system used for growing $BiMnO_3$, BLNMO and $BiFeO_3$ films. The condensed material forms an epitaxial film on the substrate if grown under the correct deposition conditions, where the film often maintains the target stoichiometry virtually unaffected. This feature may be regarded the key factor for the large success of the PLD film growth technique in recent years. Yet, enriched targets are commonly used when volatile species such as Bi are involved, where stoichiometry is often not maintained.

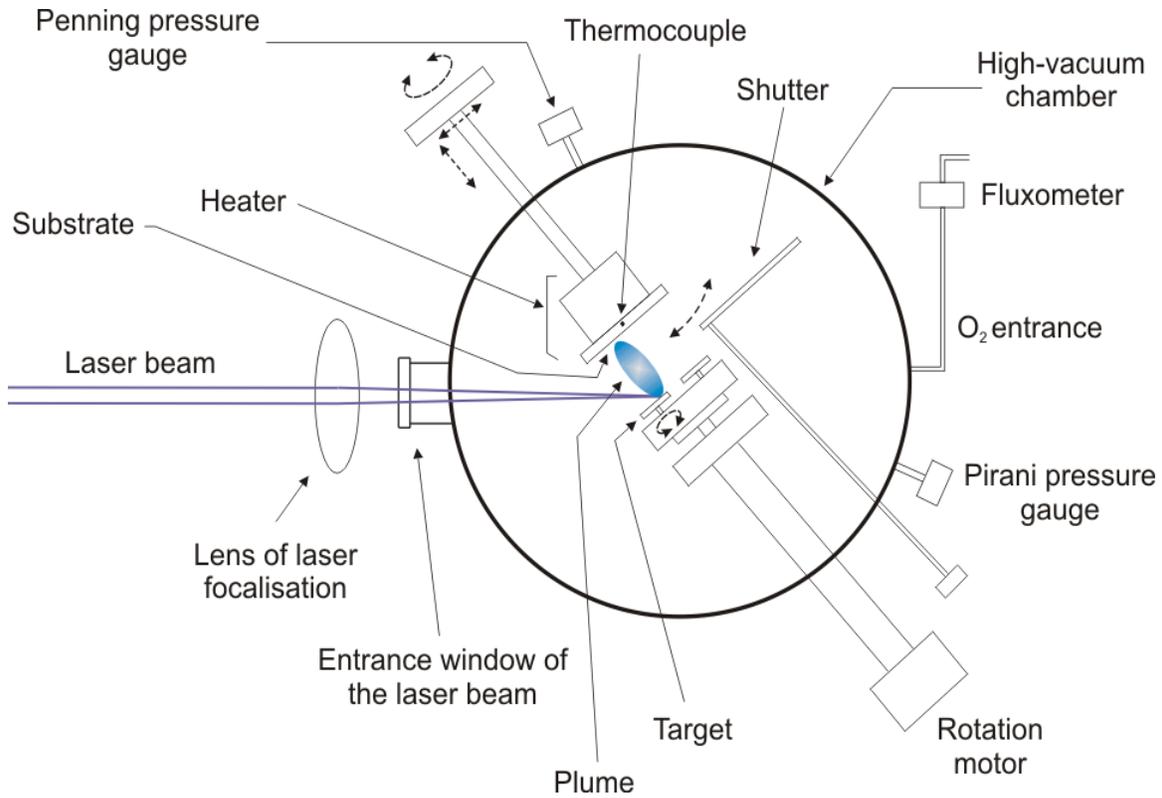

Figure 10. Schematic drawing of the PLD system used in this work.

The condensed or ablated products have higher kinetic energies (up to a few hundred eV) than in other conventional techniques such as radio frequency magnetron sputtering or thermal evaporation. This high energy of impinging particles leads to a high ad-atom surface diffusion which enables epitaxial film growth [76]. In case of oxide films, $O_2$ gas is introduced into the vacuum chamber during the growth process to an appropriate amount to prevent oxygen vacancies forming in the films. This is particularly important for the growth of insulating oxide films, where incomplete oxygenation often leads to undesired mixed cationic valence states and considerable conductivity. Furthermore, the $O_2$ pressure during deposition plays a major role on the plume expansion velocity and therefore on film growth dynamics [77], where the latter is also critically effected by the substrate temperature. The substrate selection is another issue of major importance, especially when strain engineering (i.e. replacement of mechanical pressure by epitaxial stress) is required to stabilize an otherwise unstable phase as it is the case for $BiMnO_3$ and BLNMO. In order to facilitate epitaxial growth, Nb-doped (001)-oriented $SrTiO_3$ substrates were chosen with a cubic perovskite structure and a lattice parameter of 3.905 Å as shown in Table 1.

**Table 1. Lattice parameters and mismatch ($f = a_{bulk}/a_{substrate} - 1$) with $SrTiO_3$. $a_{PC}$ stands for the pseudocubic lattice parameters in the bulk.**

| Material | Crystal Structure | Pseudo-cubic lattice parameters | Mismatch with (001) $SrTiO_3$ |
|---|---|---|---|
| $BiFeO_3$ | Rhombohedral, $R3c$ | $a_{PC}$ = 3.96 Å | $f$ = 1.41 % |
| $BiMnO_3$ | Monoclinic, $C2$ | $a \approx c$ = 3.935 Å | $f_{a,c}$ = 0.77 % |
|  |  | $b$ = 3.99 Å | $f_b$ = 2.15 % |
| $Bi_2NiMnO_6$ | Monoclinic, $C2$ | $a_{PC}$ = 3.877 Å | $f$ = -0.72 % |

In the case that epitaxial thin films are coherently grown, the in-plane lattice parameter of film and substrate are identical when misfit dislocations are absent. For $BiMnO_3$, BLNMO and $BiFeO_3$ films the following mismatch would then be expected:

- The in-plane strain in $BiFeO_3$ films would be compressive and is relatively high (1.4%), which may lead to a tendency of the $BiFeO_3$ lattice to relax with the increase of film thickness by the creation of defects.
- In $BiMnO_3$ the $a$ and $c$ lattice parameters show a lower lattice mismatch to $SrTiO_3$ (0.77%) than that of $b$ (2.15%), which suggests that it is the latter which stands out-of-plane in an epitaxial film. The in-plane strain would be compressive.
- The in-plane strain in BLNMO films would be tensile with a relatively low value of -0.71%, which may favour 2D growth.

**Table 2. Optimum deposition conditions. The laser fluence was 1.8 J/cm$^2$ with a repetition rate of 5 Hz.**

| Material | % Bi target enrichment | $O_2$ pressure (mbar) | Substrate temperature (ºC) |
|---|---|---|---|
| $BiFeO_3$ | 10 | 0.1 | 650 |
| $BiMnO_3$ | 10 | 0.1 | 630 |
| $(Bi_{0.9}La_{0.1})_2NiMnO_6$ | 0 | 0.5 | 620 |

To obtain single phase $BiMnO_3$, BLNMO and $BiFeO_3$ films all critical deposition parameters need to be adjusted such as $O_2$ pressure, substrate temperature and film thickness. For the work presented here a narrow window for the values of such parameters was encountered for all 3 species due to considerable Bi volatility as shown in Table 2 [31, 62].

# 3. RESULTS

## 3.1. Structural Characterization

### 3.1.1. X-Ray Diffraction

#### 3.1.1.1. Phase Identification and Out-of-plane Orientation

X-ray diffraction (XRD) on BiMnO$_3$, BLMNO and BiFeO$_3$ films was carried out using a PANalytical Pro MRD 4-circle diffractometer using Cu K$_\alpha$ radiation. Figure 11 shows representative θ/2θ diffractograms of BiFeO$_3$, BiMnO$_3$ and BLNMO thin films grown on SrTiO$_3$ substrates.

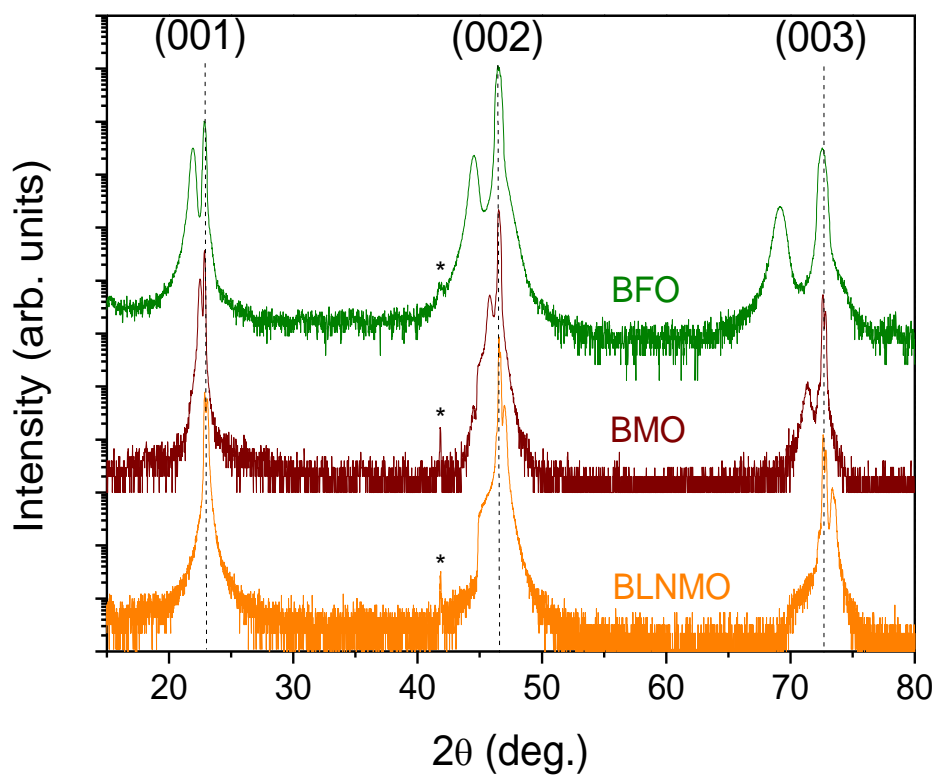

Figure 11. XRD symmetric θ-2θ diffractograms of the BiFeO$_3$ (top), BiMnO$_3$ (center) and BLNMO films (bottom). Substrate peaks are indicated by dashed lines. The asterisks (*) indicate the SrTiO$_3$ (002) peak caused by Cu K$_\beta$ radiation.

In all samples the SrTiO$_3$ (00ℓ) substrate peaks are indicated by dashed lines, accompanied by the film peaks which correspond to the (00ℓ) reflections for BLNMO and BiFeO$_3$ and to the (0k0) for BiMnO$_3$ in pseudocubic notation. No spurious phases are found in the XRD patterns.

Table 3. Out-of-plane lattice parameters obtained from θ/2θ diffractograms.

| Material | BiFeO$_3$ | BiMnO$_3$ | (Bi$_{0.9}$La$_{0.1}$)$_2$NiMnO$_6$ |
|---|---|---|---|
| Out-of-plane lattice parameter (Å) | 4.04 | 3.99 | 3.87 |

As justified in section 2, BiMnO$_3$ has (010) out-of-plane orientation, whereas BLNMO and BiFeO$_3$ are (001)-oriented. Out-of-plane lattice parameters are listed in Table 3. In comparison to bulk values (Table 1) the out-of-plane lattice parameter is expanded for BiFeO$_3$, because of the compressive strain exerted by the substrate, whereas that of BLNMO is contracted because of the tensile strain. Yet, the lattice parameter of BiMnO$_3$ remains unchanged despite the compressive strain and the expected expansion in the out-of-plane direction to preserve the unit cell volume. This might be due to the presence of some Bi vacancies, which may form due to the high volatility of Bi during film deposition. Such cation vacancies are known to produce a reduction of the unit cell volume [30, 78]. Contrary to the common assumption that epitaxial strain preserves the bulk unit cell volume, in many epitaxial films this is not accomplished [79 - 81], as seems to be the case here for BiMnO$_3$ films.

### 3.1.1.2. ϕ-Scans and In-plane Orientation

The in-plane texture and in-plane orientation of the unit cells were characterized by ϕ-scans and pole figures of the (204) asymmetric reflection representing a lattice plane that is not parallel to the interface (Figure 12). Such reflection proved to be appropriate to discern between the film and substrate reflections, which is critical due to the similarity of the lattice parameters. A higher 2θ reflection is useful in such case, because the difference in 2θ for substrate and film is larger the higher the 2θ angle is.

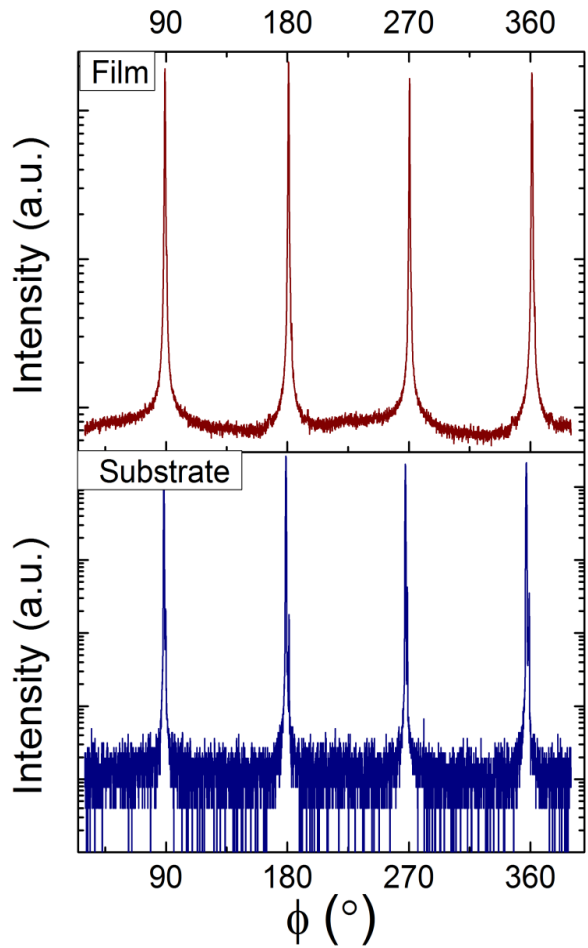

Figure 12. XRD ϕ-scans around the (204) reflection of both, 125-nm BLNMO film and SrTiO$_3$ substrate.

Figure 12 depicts the XRD ϕ-scans of the assymetric (204) reflection of both substrate and film of a 125-nm BLNMO film. Four 90º- ϕ-spaced peaks of the film are observed, which indicates that contributions from misoriented planes are absent. This, in turn, implies that BLNMO films are coherently in-plane oriented. In order to confirm the in-plane texture, pole figures were obtained where ψ was scanned from 0 to 30º, as illustrated in Figure 13 for a 60-nm BLNMO film.

In agreement with the ϕ-scans, Figure 13 reproduces the four 90º-ϕ-spaced peaks of the film, confirming the in-plane four-fold orthogonal symmetry. Moreover, the fact that the values of the ϕ-peaks coincide with the ϕ-angles of the substrate confirms that BLNMO films have grown in a cube-on-cube fashion. In-plane film alignment with substrate and hence cube-on-cube epitaxy was also confirmed for $BiFeO_3$ and $BiMnO_3$ in the same way.

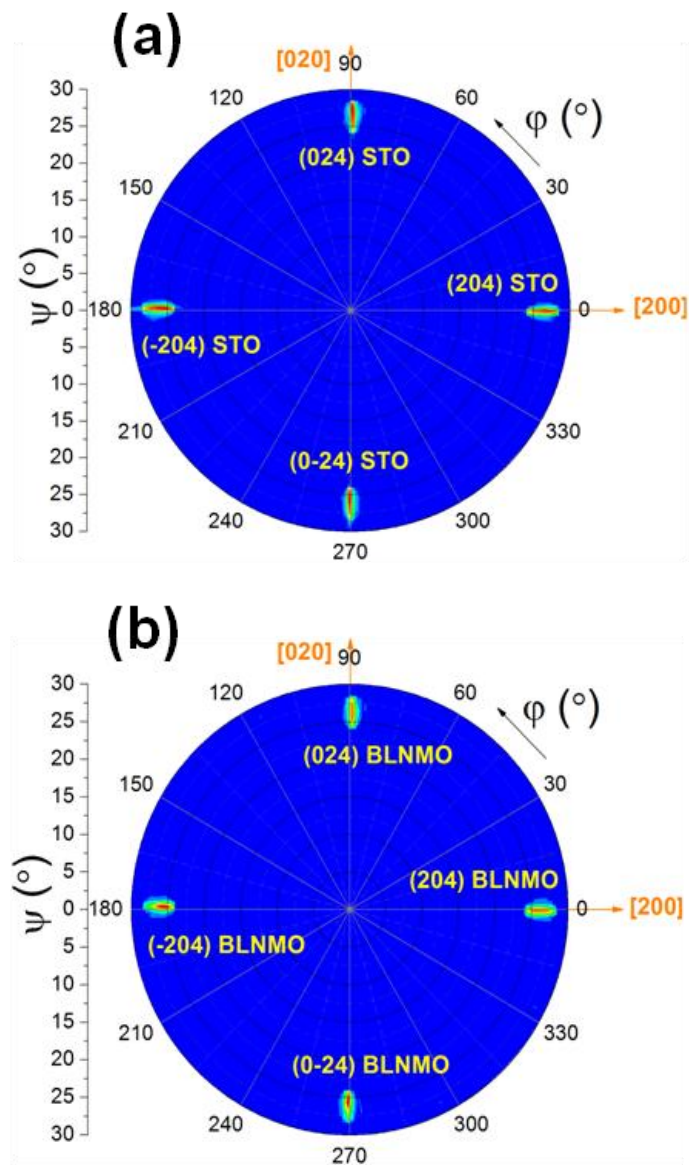

Figure 13. XRD pole figures around the (204) reflection of (a) $SrTiO_3$ substrate, and (b) 60-nm BLNMO film.

### 3.1.1.3. Reciprocal Space Maps

In reciprocal space maps (RSMs) the diffraction vector $\vec{q}$ is not directly measured, but several coupled $2\theta/\omega$ XRD scans are performed for different $\omega$ values, giving rise to a 2D $2\theta$ - $\omega$ map of intensities, and $q_{perp}$ and $q_{parallel}$ are obtained via trigonometric relationships. Figure 14 shows the RSMs around the (204) reflection for (a) $BiFeO_3$, (b) $BiMnO_3$, and (c) BLNMO. Both, the $SrTiO_3$ and the film reflections are displayed in each case, where $q_{parallel}$ of (204) is found to coincident with the substrate as indicated by the dashed lines.

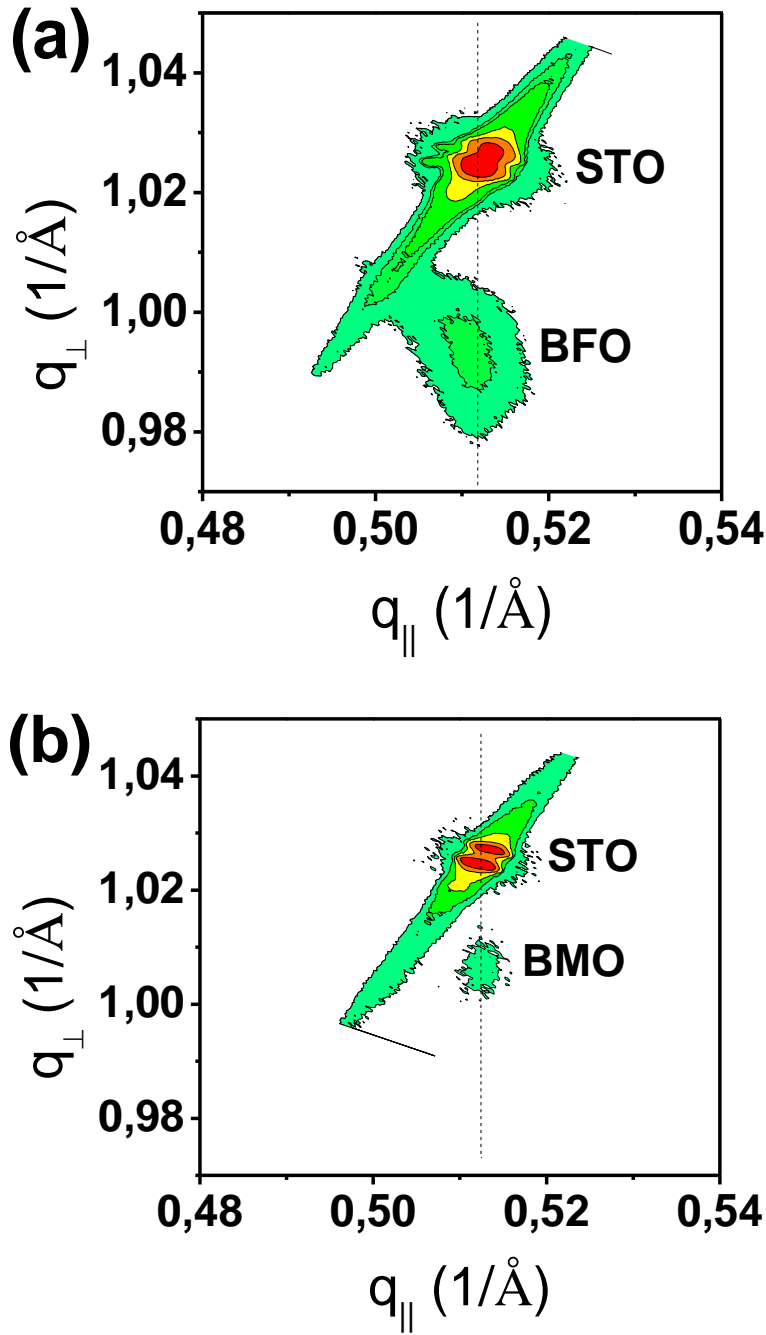

*Figure 14 to be continued on next page.*

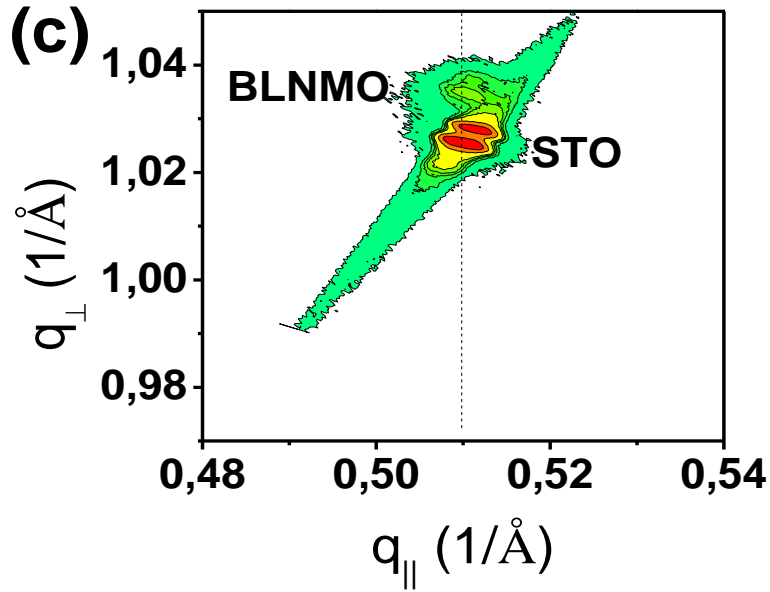

Figure 14. XRD reciprocal space maps of (a) $BiFeO_3$, (b) $BiMnO_3$ and (c) BLNMO around the (204) reflection of $SrTiO_3$. The dashed lines are a guide to the eye, indicating the in-plane $q$-value of the substrate reflection.

All samples adopt the in-plane lattice parameter of $SrTiO_3$ and, therefore, the films are fully strained. On the other hand, the high $BiFeO_3$ lattice mismatch (see Table 1) leads to a progressive relaxation of elastic energy, which is illustrated in form of a peak tail in Figure 14a extending towards lower $q_{par}$ (higher in-plane lattice parameter) and higher $q_{perp}$ (lower out-of-plane lattice parameter). The out-of-plane lattice parameters obtained from RSMs were coincident with those obtained from θ/2θ measurements (Table 2) as expected.

### 3.1.1.4. B-Site Order in $(Bi_{0.9}La_{0.1})_2NiMnO_6$ Detected by Synchrotron XRD

For a long-range ferromagnetic interaction in BLNMO films, Ni and Mn cations are required to be located alternatingly along the [104], [010] and [001] crystallographic directions, which results in the alternation of Ni-planes and Mn-planes along the [111] direction (along the eight equivalent directions: [1-11], [11-1], [-111], [1-1-1], [-11-1], [-1-11], [-1-1-1], [111]). As a consequence, the occurrence of the (1/2 1/2 1/2) superstructure XRD reflection should be expected. However, as the atomic scattering factor of $Ni^{2+}$ and $Mn^{4+}$ are very similar, the superstructure XRD reflection in $Bi_2NiMnO_6$ is very weak. Therefore, synchrotron radiation XRD was used, which provides an intense source of X-rays for detecting weak XRD reflections.

Figure 15a and b depict θ/2θ synchrotron XRD scans around (111) and (1/2 1/2 1/2) BLNMO reflections respectively, for a 100-nm BLNMO film. Due to the similar lattice parameters of BLNMO and $SrTiO_3$, the perovskite (111) reflection of both film and substrate ($2θ_{film} \approx 32.62°$ and $2θ_{susbtrate} \approx 32.59°$, respectively) can be seen in Figure 15a. Contrarily, as superstructure order is absent in $SrTiO_3$, the peak at $2θ \approx 16.14°$ observed in Figure 15b solely corresponds to the BLNMO superstructure (1/2 1/2 1/2) reflection. Therefore, long-range B-site cation order in BLNMO thin films was confirmed unequivocally. The equivalent superstructure peaks were also observed on selected area electron diffraction patterns using high resolution electron microscopy [58].

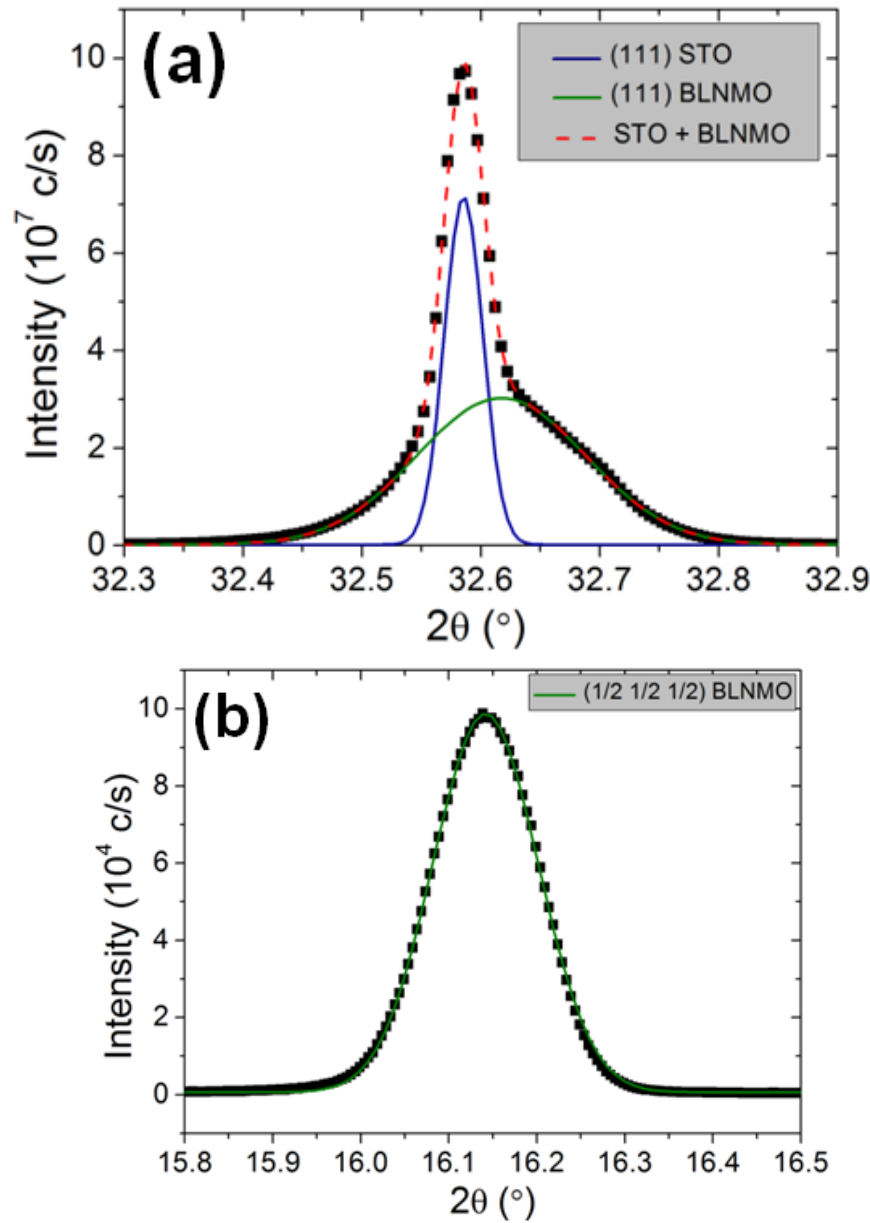

Figure 15. (a) θ/2θ synchrotron XRD scans around the perovskite (111) reflection for BLNMO and SrTiO$_3$. (b) θ/2θ synchrotron XRD scan around the (1/2 1/2 1/2) superstructure reflection for BLNMO. The diffraction intensity peaks were fitted using Gaussian functions. Reproduced from Ref. [58] with permission from the American Institute of Physics (AIP).

### 3.1.2. HRTEM Microstructure Determination of $(Bi_{0.9}La_{0.1})_2NiMnO_6$

High-resolution TEM (HRTEM) images from BLNMO films are presented in Figure 16a and b. The high crystalline film quality indicated by X-rays is confirmed and no misfit dislocations are observed, as expected for a coherently strained film. The BLNMO/SrTiO$_3$ interface appears strikingly sharp with no cationic inter-diffusion across the interface. Fast Fourier Transmforms (FFTs) of selected areas of the HRTEM images were computed in order to assess the periodicity of the crystal lattice. According to the FFTs depicted in Figure 17b and c the crystallographic directions of both substrate and film are well aligned in agreement with XRD measurements.

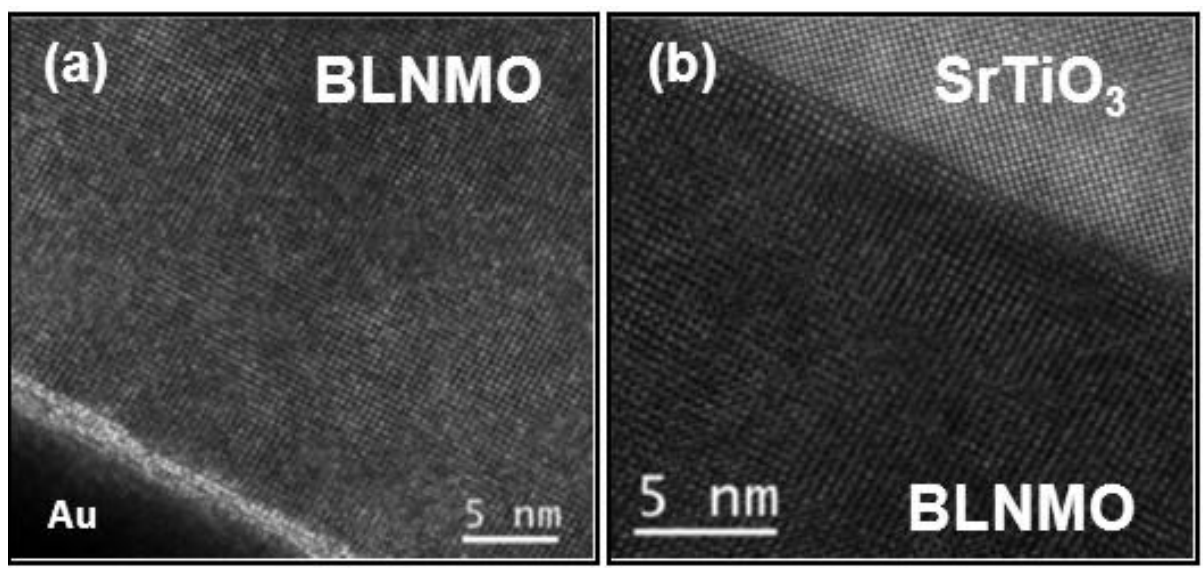

Figure 16. (a) HRTEM image of BLNMO film in cross-section geometry along the [104] direction. (b) Higher magnification HRTEM image of the BLNMO/SrTiO$_3$ interface. Reproduced from Ref. [58] with permission from the American Institute of Physics (AIP).

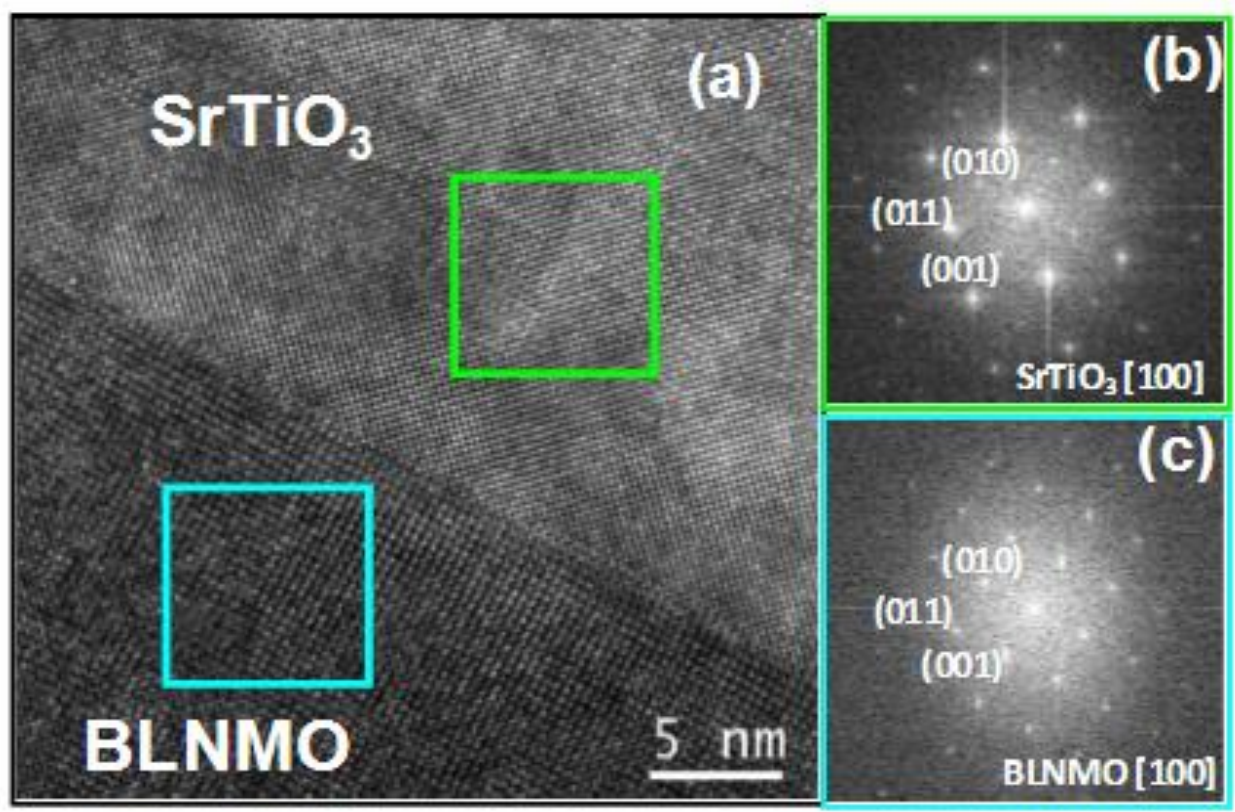

Figure 17. (a) HRTEM image of the BLNMO film and the SrTiO$_3$ substrate in cross-section geometry along the [104] direction. The green and blue squares mark the selected areas of the substrate and film where FFTs were taken. FFTs of (b) the SrTiO$_3$ substrate, and (c) of the BLNMO film. Reproduced from Ref. [58] with permission from the American Institute of Physics (AIP).

*3.1.3. Surface Topography*

The surface topography of the films was studied by atomic force microscopy (AFM) using an Agilent 5420 instrument, as presented in Figure 18.

Both $BiMnO_3$ and $BiFeO_3$ films developed faceted crystallites, evidencing 3D growth mode. Such facets can be deduced to represent coherently aligned crystal planes when studying the distribution of angles in the AFM data. Perfect alignment of these objects corroborates single-domain epitaxial growth and, as deduced from XRD analysis, their orientation coincides with the STO crystallographic axis. The root-mean-square roughness for $BiMnO_3$ and $BiFeO_3$ films was in the order of few nanometers not exceeding 5 % of the nominal film thickness.

On the contrary, the BLNMO film exhibits a quite different morphology where substrate steps and terraces are maintained. This is indicative of a 2D growth mode, in agreement with the higher epitaxial quality. Meandering can be observed as a consequence of anisotropic diffusion due to Ehrlich-Schwoebel terrace potentials [82].

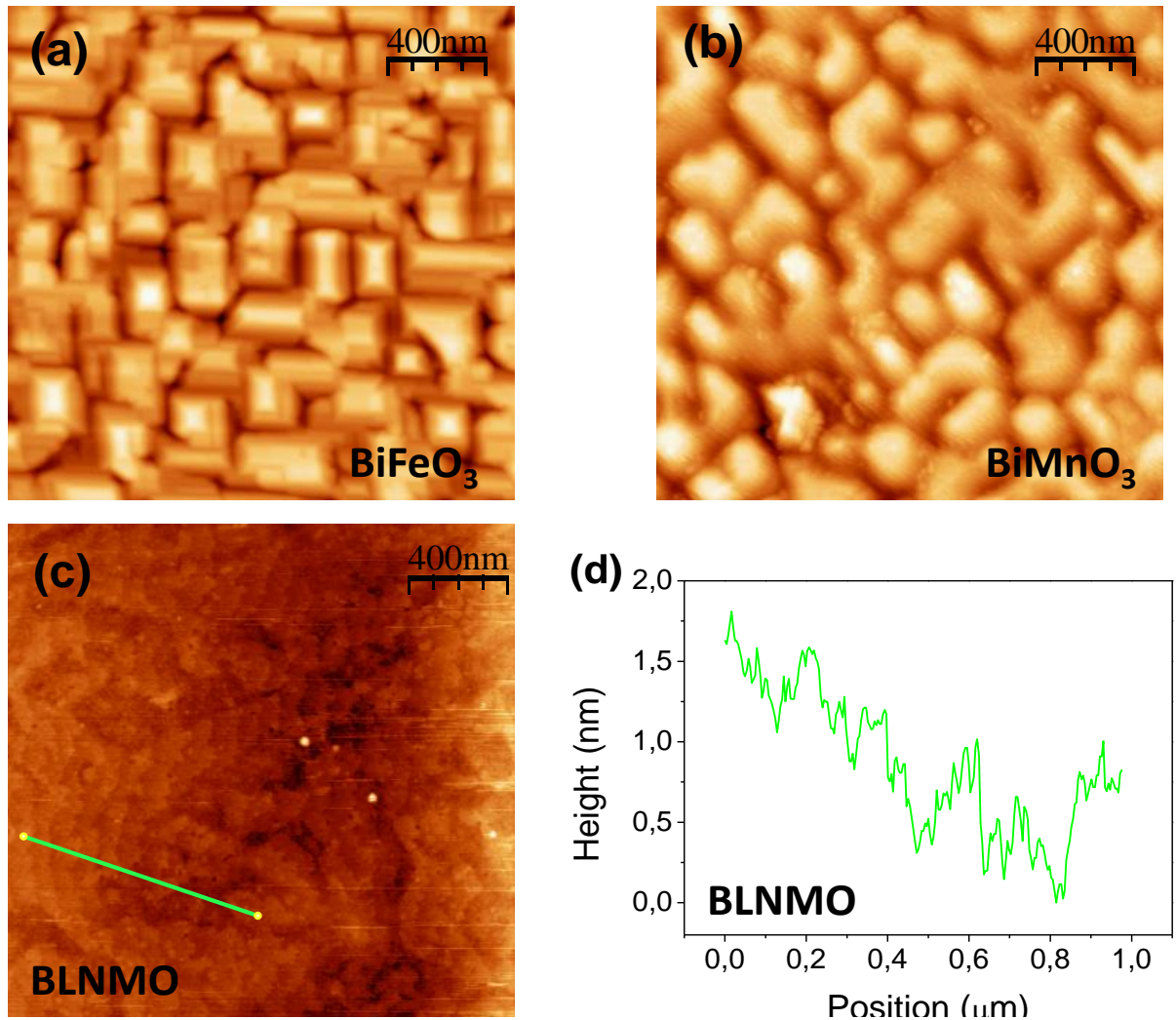

Figure 18. 2x2 μm² AFM topography images of (a) $BiFeO_3$, (b) $BiMnO_3$ and (c) BLNMO. (d) corresponds to the BLNMO film height profile indicated by the solid (green) line in (c).

## 3.2. Magnetic Properties

The magnetic properties of $BiMnO_3$, BLNMO and $BiFeO_3$ films were determined using SQUID magnetometry (Quantum Design MPMS XL) with magnetic field ($H$) applied in the film in-plane direction and the resulting film magnetization ($M$) being recorded.

### 3.2.1. BiMnO₃

The ferromagnetic moment in $BiMnO_3$ thin films is clearly observed in the hysteretic magnetization ($M$) vs applied magnetic field ($H$) curve recorded at 5 K (Figure 19). A saturated $BiMnO_3$ moment of $M_{sat.} \approx 2.5$ $\mu_B$/f.u. was found, which is smaller than the bulk value (3.6 $\mu_B$/f.u.) but close to the thin film value reported previously (2.2 $\mu_B$/f.u.) [27]. A $BiMnO_3$ coercive field $H_{coerc.} \approx 425$ Oe (upper inset of Figure 19) was found. The temperature dependence of $M$ under 1 kOe applied $H$ and the differentiated inverse of the magnetic susceptibility $1/\chi$ (lower inset of Figure 19) allowed determining the ferromagnetic Curie temperature $T_C \approx 100$ K, in close agreement to the bulk value [83-85].

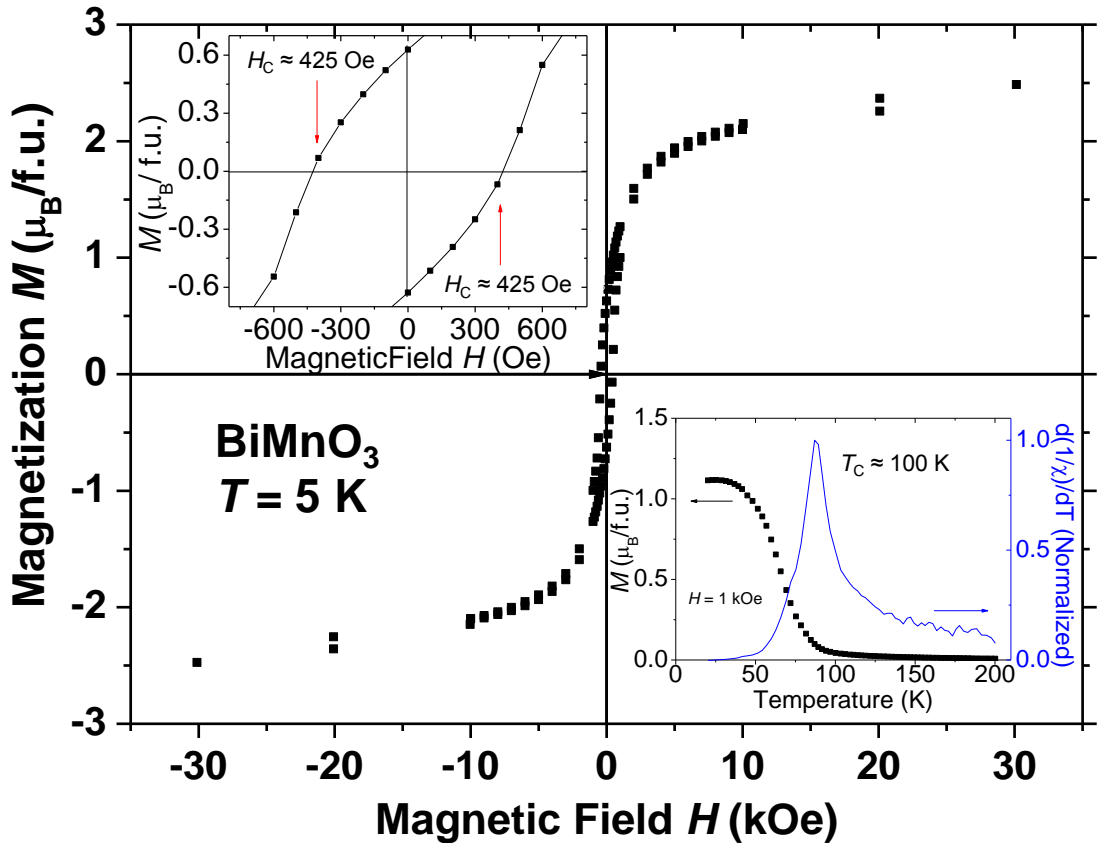

Figure 19. Magnetic field ($H$) dependence of magnetization ($M$) for a $BiMnO_3$ thin film at 5 K. Upper inset: Zoom of the low-field region. Lower inset: $T$ dependence of $M$ under field-cooled (FC) conditions using an external in-plane field of $H$ = 1 kOe (left axis) and derivative of the inverse susceptibility (right axis).

### 3.2.2. (Bi₀.₉La₀.₁)₂NiMnO₆

BLNMO films are B-site ordered, fully stoichiometric and the electronic configuration of $Ni^{2+}$ and $Mn^{4+}$ cations are those required for ferromagnetic coupling [58]. An ideal ferromagnetic ordering of the spins of the B-cations, $Ni^{2+}$ ($t_{2g}^6 e_g^2$, S=1) and $Mn^{4+}$ ($t_{2g}^3$, S=3/2), implies a saturated magnetization of $M_S = 5$ $\mu_B$/f.u.

Magnetization data of Figure 20 shows that the film does not saturate even at a field of 7 T, although the diamagnetic substrate contribution has been subtracted from the data. The films have a magnetization $M_S$ (7 T, 10 K) ≈ 3.5 $\mu_B$/f.u., in close agreement with data reported for bulk $(Bi_{0.9}La_{0.1})_2NiMnO_6$ samples ($M_S$ = 3.6 $\mu_B$/f.u.), which do not saturate either at 7 T field and 5K [65]. The presence of antisite defects and antiphase boundaries, common defects in double perovskite structures [86], may account for both the hardness to saturation and the accompanying reduced magnetization.

The remanent magnetisation, as shown in Figure 20 (upper panel), is rather low probably due to the small spin-orbit coupling of the magnetic cations.

On the other hand, inspection of the temperature dependent magnetization data (lower panel in Figure 20) indicates that the Curie transition occurs at $T_C$ ≈ 100 K. This value is lower than in the bulk (≈ 140 K) [59], but this is in agreement with the reduced $T_C$ found previously for BLNMO thin films [60, 61]. The reduction in the magnetic transition temperature has been ascribed to the epitaxial stress exerted by the substrate [60, 62]. The rather broad transition temperature may result from a distribution of superexchange interactions within the sample, likely driven by a dispersion of the distinct length and angles of Ni - O - Mn bonds near the A-sites, which are occupied by either La or Bi cations.

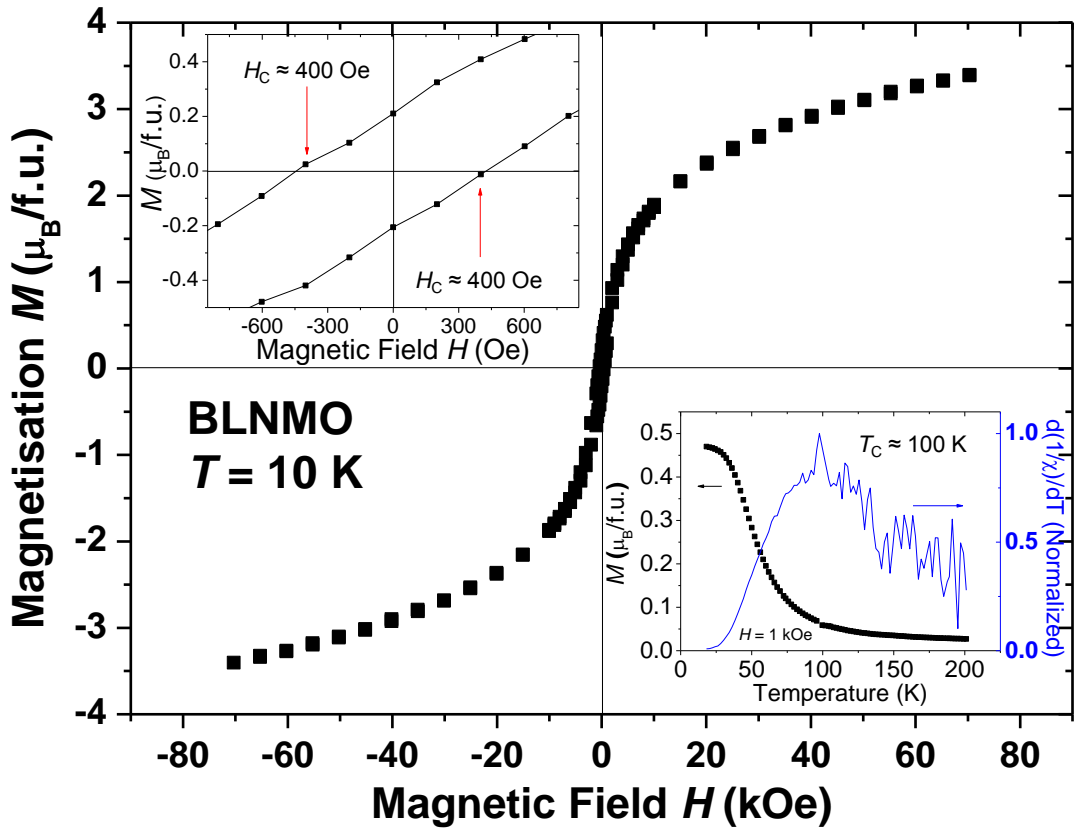

Figure 20. Magnetic field ($H$) dependence of magnetization ($M$) for a BLNMO thin film at 10 K. Upper inset: Zoom of the low-field region. Lower inset: $T$ dependence of $M$ under field-cooled (FC) conditions using an external in-plane field of $H$ = 1 kOe (left axis) and derivative of the inverse susceptibility (right axis).

### 3.2.3. BiFeO₃

The $BiFeO_3$ film showed signs of antiferromagnetism with a weak ferromagnetic moment up to room temperature and the $M_{sat.}$ ≈ 0.05 $\mu_B$/f.u. registered at 5 K (Figure 21) agrees with previous reports [45, 87-89].

The appearance of a small ferrimagnetic component in epitaxial $BiFeO_3$ thin films may be ascribed to spin canting due to lattice or FE domain wall strain or oxygen vacancies [89].

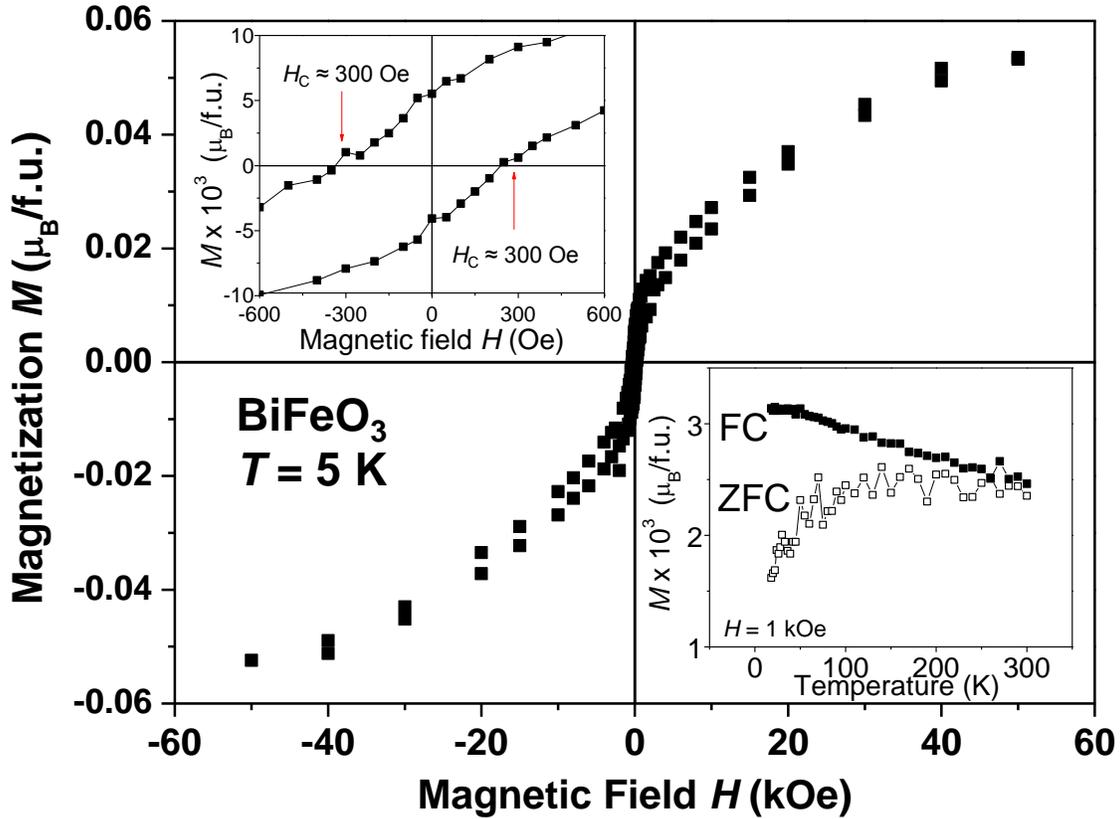

Figure 21. $H$ dependence of $M$ for $BiFeO_3$ thin films at 5 K. Upper inset: Zoom of the low-field region. Lower inset: Zero-field cooled (ZFC) and field cooled (FC) $M$ vs $T$ curves using an external in-plane magnetic field of 100 Oe.

## 3.3. Ferroelectric Properties

As mentioned in section 1, ferroelectricity in Bi containing perovskites commonly arises from the off-centering of the lone-pair electrons of Bi cations. If the displacement vectors of the lobe-distributed $Bi^{3+}$ lone-pair electrons align spontaneously and coherently on a macroscopic scale, the conditions for ferroelectricity are fulfilled and ferroelectric domains form, which may be poled or switched by the application of sufficiently high electric fields.

### 3.3.1. Ferroelectric Hysteresis Loop Measurements

The most commonly used technique to detect ferroelectricity may be ferroelectric polarization measurements by using a time-dependent sinusoidal electric field to detect hysteretic behavior in form of a hysteresis loop. Since the poling or switching of ferroelectric domains by electric fields implies that $Bi^{3+}$ lone-pair electrons change their lattice position, such electron movements can be picked up at the electrodes as a small current pulse at the moment when the switching occurs. Such ferroelectric testing can be ambiguous though [90], because applied electric fields may also cause leakage currents which can mask the switching current. Such leakage currents can be determined and subtracted from the ferroelectric switching, which then may lead to hysteretic curves consistent with ferroelectric polarization.

*3.3.2. Piezo-force Microscopy*

A more reliable technique to test for the switching of ferroelectric domains may be piezo-force microscopy, where ferroelectric domains are switched locally by applying a voltage between a scanning-probe microscopy (SPM) tip and the sample. The local switching can be monitored by the piezo-force response of the ferroelectric sample. This is demonstrated on 100 nm $BiFeO_3$ thin films deposited on conducting Nb-STO substrates as shown in Figure 22.

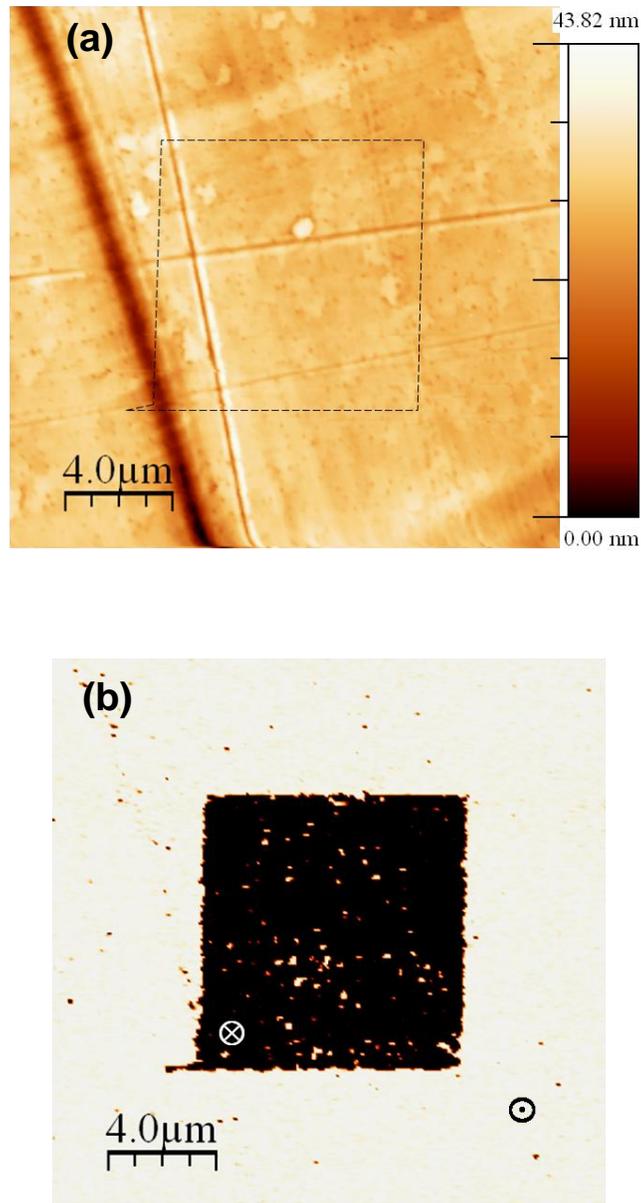

*Figure 22 to be continued on next page.*

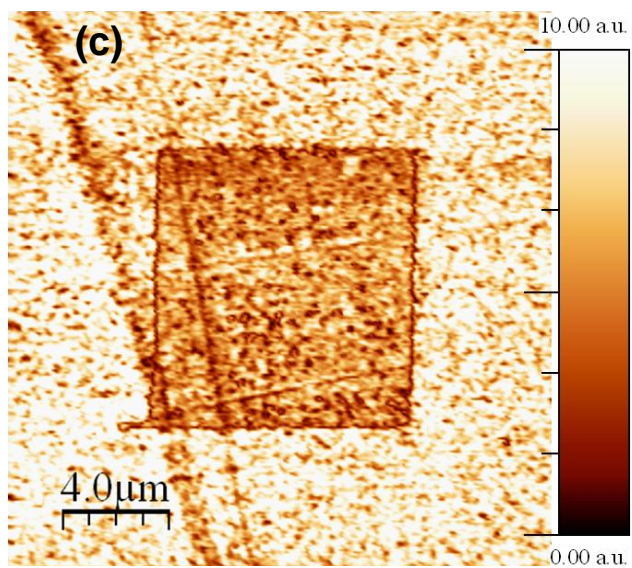

Figure 22. (a) Surface topography, (b) out-of-plane PFM phase, and (c) out-of-plane PFM amplitude images (20 x 20 µm$^2$) for a BiFeO$_3$ thin film at room temperature. The marked area in (a) was previously scanned by applying a DC voltage between tip and substrate (-5 V).

Images were acquired using Si tips with PtIr coating at an AC voltage of 0.5 Vrms and a frequency of 50 kHz. In the set-up used here voltages were applied to the sample while the tip was grounded. Well-defined poled areas were patterned into the film by applying a DC voltage between tip and substrate, and subsequently checked by piezo-response force microscopy (PFM). Simultaneous topographic, out-of-plane PFM phase and amplitude images (Figures 22a, b, c) were recorded.

The topographic image in Figure 22a shows that switching the polarization within the marked area does not damage the BiFeO$_3$ film surface. The phase image in Figure 22b presents a 180º contrast between the poled area and the surrounding film, pointing to the switching of the perpendicular component of the polarization, with a corresponding minimum in the amplitude image at the domain wall (Figure 22c). Polarization can be switched back and forth by reversing the polarity of the applied voltage, confirming that the BiFeO$_3$ film was ferroelectric at room temperature. PFM experiments on BiMnO$_3$ and BLNMO films at room temperature are not feasible due to low film resistance.

A further possibility to detect for ferroelectricity is the use of dielectric spectroscopy to measure the dielectric permittivity $\varepsilon'$ at various temperatures. In ferroelectric materials $\varepsilon'$ vs $T$ curves exhibit a maximum at the transition temperature $T_C$. However, these measurements can be troublesome in the presence of multiple dielectric relaxations and the direct observation of switching ferroelectric domains may always be preferable.

## 3.4. Magneto-electric Coupling Detected by Magnetic Field Dependent Impedance Spectroscopy

For potential industrial application of Bi containing multiferroic perovskites, a strong coupling between magnetic and electric order parameters is desired. Such magneto-electric coupling (MEC) effects are difficult to be detected, which is a major problem in the emerging field of multiferroics. Sophisticated techniques such as transmittance optical spectroscopy [91], X-ray absorption [92] or scattering [93] and magnetic field dependent PFM [94] are expected to yield unambiguous results. On the other hand, much discussion has arisen over the more frequently used technique of dielectric spectroscopy, which in multiferroics research is often combined with applied magnetic fields.

The use of such frequency ($f$), temperature ($T$) and magnetic field ($H$) dependent impedance spectroscopy (IS) allows detection of and discrimination between magneto-resistance (MR) and magneto-capacitance (MC) effects. It has been pointed out previously that such discrimination is a non-trivial task, because pure MR of a secondary and extrinsic dielectric relaxation in the sample can cause changes in capacitance [95, 96]. Therefore, it has been argued that all existing dielectric relaxations within one sample need to be identified and deconvoluted first, before a magnetic field can be applied. In polycrystals multiple dielectric relaxations may originate from grain interior (bulk) regions, grain boundaries (GBs) and electrode-sample interface (IF) layers [97], whereas in epitaxial thin films only bulk (or intrinsic film) and IF contributions are expected [98]. The MC and MR of each contribution in the sample must be determined separately by IS for a range of applied magnetic fields. In this way the true MC and MR of all separated relaxations can be obtained. Such analysis may then be repeated at various $T$ to obtain a comprehensive picture of the intrinsic MC and MR behavior. This procedure is demonstrated here, where first all dielectric relaxations occurring in the samples are identified and deconvoluted.

Impedance spectroscopy measurements on all films were carried out in the out-of-plane direction, which leads to reasonable thin film resistance in nominally insulating materials. For such purpose all films were deposited on highly conducting Nb-doped $SrTiO_3$ substrates, where the film surfaces would be covered with sets of Pt electrodes. In such way the impedance can be measured twice across the film, which was shown previously to be appropriate to obtain valid IS data from epitaxial multiferroic thin films [98].

### 3.4.1. Deconvolution of Dielectric Relaxations

IS data are usually represented on plots of the imaginary vs the real part of the impedance ($-Z''$ vs $Z'$) and for each series dielectric relaxation ideally one semicircle appears in agreement with the RC element model, where each series relaxation is represented by a parallel resistor - capacitor (RC) element [97, 99, 100]. In multiferroics research however, dielectric data are often better represented in terms of dielectric permittivity $\varepsilon'$ vs $f$ or $\varepsilon'$ vs $T$ to directly display changes of $\varepsilon'$ as a result of applied $H$, or $\varepsilon'$ changes with $T$ across a magnetic transition.

Impedance spectroscopy data from 50 nm $BiMnO_3$, 100 nm BLNMO and 100 nm $BiFeO_3$ thin films are represented in the preferable $\varepsilon'$ vs $f$ notation in Figure 23a, b and c respectively. Clear signs of two dielectric relaxation processes were obvious in $BiMnO_3$ and BLNMO films as manifested by the existence of two $\varepsilon'$ plateaus, $\varepsilon'_{high}$ and $\varepsilon'_{low}$. Such plateaus corresponds to two dielectric relaxations in series as represented by a series of two RC elements, where the low permittivity plateau $\varepsilon'_{low}$ represents the intrinsic $BiMnO_3$/BLNMO film permittivity and $\varepsilon'_{high}$ an extrinsic Maxwell-Wagner type electrode interface relaxation [96-99]. The $BiMnO_3$ $\varepsilon'_{low}$ plateau was found to be ≈ 35 - 40, in a similar range as $BiMnO_3$ permittivity values of ≈ 30 reported before [22, 98, 101]. For BLNMO, $\varepsilon'_{low}$ ≈ 100 was found to be larger as compared to $BiMnO_3$. In the $BiFeO_3$ film no $\varepsilon'$ plateaus are visible and in the $T$-range investigated only the intrinsic film contribution $\varepsilon'_{low}$ is visible, displaying values in the range of ≈ 90 - 120. The strong $f$ dependence of the intrinsic $BiFeO_3$ capacitance is a sign of a certain degree of non-ideality of the dielectric relaxation [104], which is also visible for the $BiMnO_3$ and BLNMO relaxations, where the $\varepsilon'_{high}$ and $\varepsilon'_{low}$ plateaus both show a slight $f$ dependence too.

The $BiMnO_3$ and BLNMO data were fitted to an equivalent circuit of a series of two RC elements (see inset in Figures 22a and b), whereas the single relaxation in $BiFeO_3$ was fitted by one RC element. In all spectra for $BiMnO_3$, BLNMO and $BiFeO_3$, extrinsic parasitic contributions were accounted for by a series inductor L0 describing the inductance of the measurement wires and a series resistor R0 describes the resistance of the conducting substrate, the electrodes and measurement wires [98]. The non-ideality of the intrinsic $BiMnO_3$ and BLNMO film relaxations was accounted for by a parallel constant-phase element (CPE1) and for the extrinsic $BiMnO_3$/BLNMO relaxations by the replacement of the ideal capacitor by CPE2.

Bi Containing Multiferroic Perovskite Oxide Thin Films 29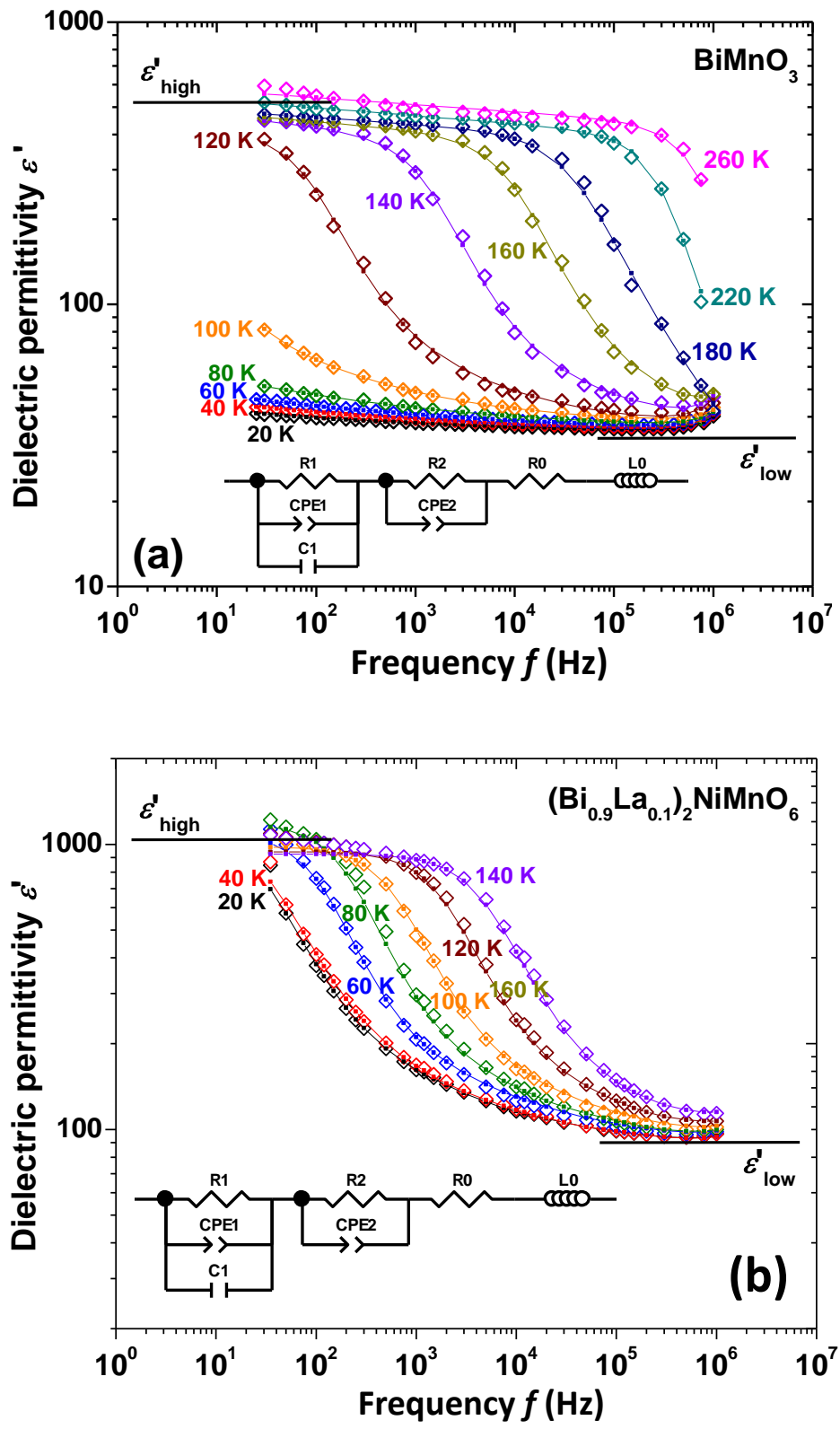

*Figure 23 to be continued on next page.*

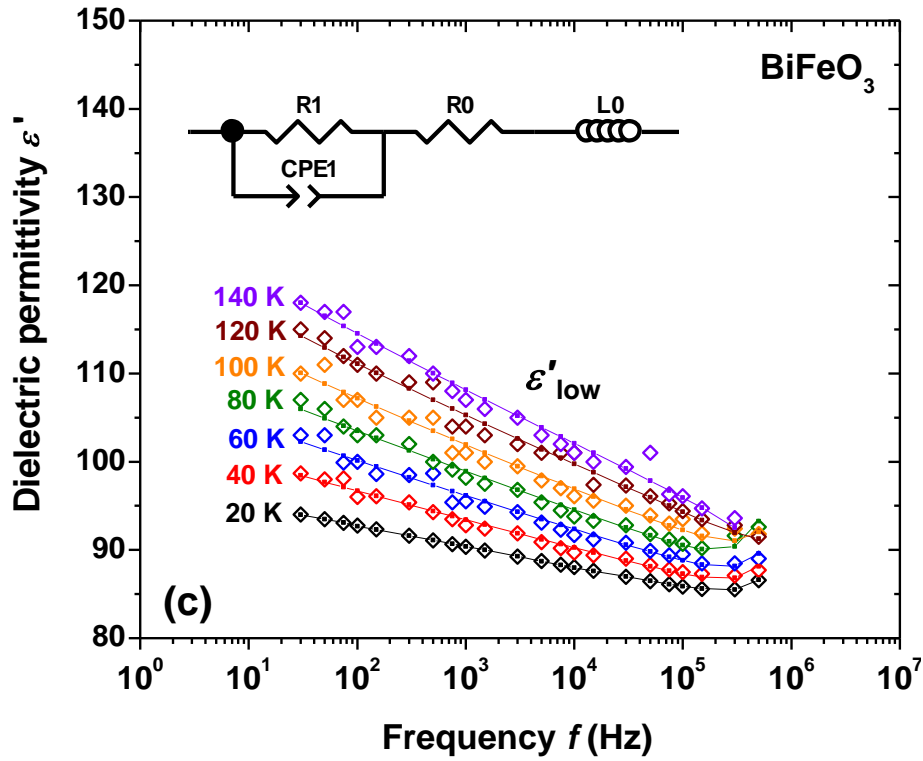

Figure 23. Dielectric permittivity $\varepsilon'$ vs $f$ for (a) 50 nm BiMnO$_3$, (b) 100 nm BLNMO, and (c) 100 nm BiFeO$_3$ films at selected temperatures as indicated. Open symbols (◊) represent experimental data, full squares (■) and solid lines represent fits to the data using the equivalent circuit depicted as figure insets. The occurrence of two $\varepsilon'$ plateaus ($\varepsilon'_{high}$ and $\varepsilon'_{low}$) for BiMnO$_3$ and BLNMO is consistent with two dielectric relaxations in series, as represented by two series RC elements. The presence of one $\varepsilon'$ plateau ($\varepsilon'_{low}$), or better a constant $\varepsilon'$ vs $f$ slope, for BiFeO$_3$ is consistent with one dielectric relaxation, as represented by one RC element. R0 and L0 represent parasitic contributions.

Such replacement of the ideal capacitor also gave good fits to the intrinsic BiFeO$_3$ film relaxation in terms of the R1-CPE1 element depicted in Figure 23c. The CPE carries a critical exponent $n$ [99], which has typical values of $n = 0.6 - 1$. $n = 1$ constitutes the case of an ideal capacitor for an ideal dielectric relaxation. In a non-ideal $R$-CPE circuit decreasing $n$ values indicate a broadening of the respective dielectric relaxation peak as a reflection of the broadening of the distribution of relaxation times, $\tau$, across the sample. In an ideal RC element $\tau$ is given by $\tau = RC$. The exact shape of the distribution of $\tau$ is complicated to be determined from IS data, and the exponent $n$ constitutes a semi-empirical parameter to reflect an increasing width of the distribution in $\tau$ by decreasing $n$.

The fact that in BiMnO$_3$ and BLNMO the intrinsic film relaxations could only be fitted by a R1-CPE1-C1 element and not by a more simple R1-CPE1 element is debited to the fact that the $\varepsilon'$ drop from $\varepsilon'_{high}$ to $\varepsilon'_{low}$ in $\varepsilon'$ vs $f$ is not as sharp as expected for an ideal intrinsic relaxation. Additionally, the $\varepsilon'_{high}$ and $\varepsilon'_{low}$ plateaus both exhibit a slight $f$-dependence. Such multiple signs of non-ideality can only be fitted with an R1-CPE1-C1 element for the intrinsic and an R2-CPE2 element for the extrinsic contribution.

In order to confirm the validity of each equivalent circuit presented in Figure 23a, b and c, and to get a clearer picture of the physics behind each deconvoluted dielectric relaxation it is advantageous to consider the temperature trends of all parameters extracted from the respective equivalent circuit components.

### 3.4.2. Equivalent Circuit Fitted Parameters for BiMnO$_3$, BLNMO and BiFeO$_3$ Films

Using the equivalent circuits shown in the insets of Figure 23a, b and c the BiMnO$_3$, BLNMO and BiFeO$_3$ intrinsic and extrinsic relative dielectric permittivity, $\varepsilon_1$ (C1) and $\varepsilon_2$ (C2) respectively, the intrinsic and extrinsic resistivity, $\rho_1$ (R1) and $\rho_2$ (R2) respectively, and the CPE exponents $n$ were extracted from the fits at various $T$. The resulting $T$ dependences of such deconvoluted parameters are presented in Figures 24 and 25, where each data point represents the fitted value of the respective circuit component at the respective temperature.

The BiFeO$_3$ intrinsic film permittivity $\varepsilon_1$ increases uniformly from 100 to 130 with $T$ (Figure 24a), which is a clear sign for ferroelectricity. The intrinsic BiMnO$_3$ permittivity $\varepsilon_1 \approx 35 - 45$ shows a modest increase up to 75 K, again consistent with ferroelectricity. At higher $T$ around the magnetic transition $T_C \approx 100$ K, the BiMnO$_3$ intrinsic $\varepsilon_1$ vs $T$ curve displays a peak structure as demonstrated in the enlarged inset of Figure 24b. This may well represent an intrinsic MEC response of $\varepsilon_1$ to the BiMnO$_3$ magnetic transition at $T_C \approx 100$ K, since all extrinsic contributions had been deconvoluted and were accounted for by means of the equivalent circuit. This peak in $\varepsilon_1$ vs $T$ is consistent with previous work [101], where the peak was stronger pronounced though. Since the peak structure observed here is rather weak, it can hardly be resolved from the experimental data in Figure 23a and only emerges from fitting the $f$-dependent dielectric data to the equivalent circuit model.

Surprisingly, the intrinsic BLNMO film permittivity $\varepsilon_1$ did not show any signs of a peak structure in the $\varepsilon_1$ vs $T$ curve near the magnetic transition at $T_C \approx 100$ K and MEC coupling was not indicated. Obviously, the interruption of the Mn sublattice by Ni cations leads to an attenuation or disappearance of the MEC effect.

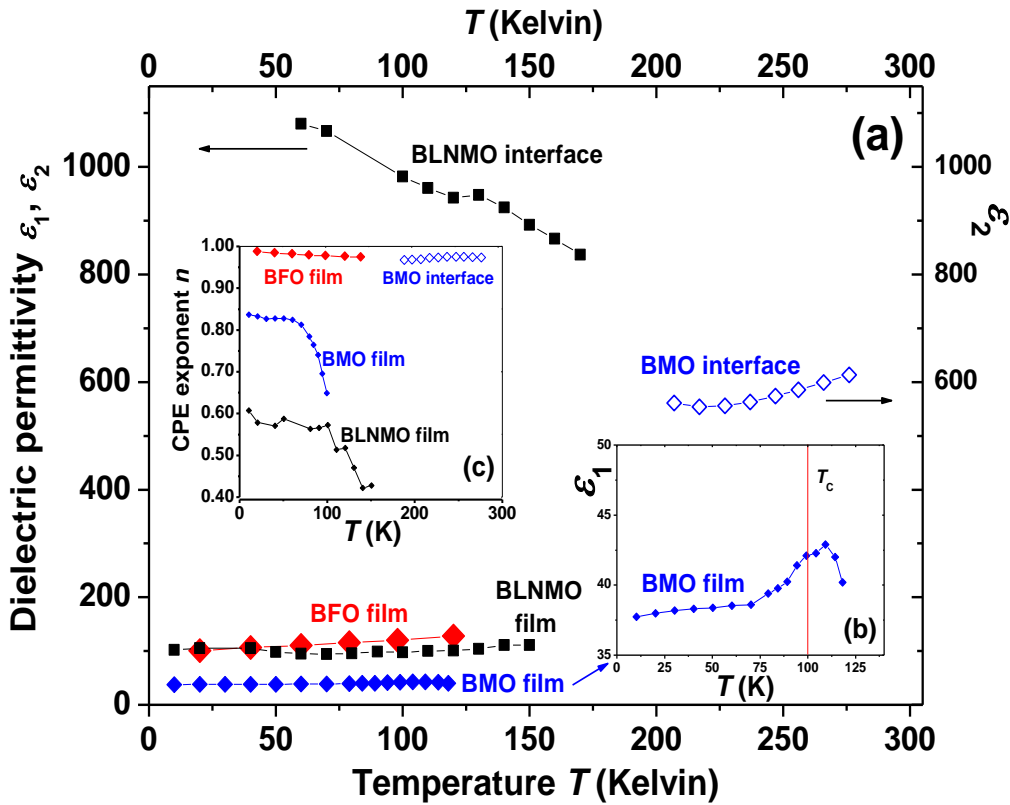

Figure 24. (a) BiMnO$_3$, BLNMO and BiFeO$_3$ intrinsic and extrinsic dielectric permittivity $\varepsilon_1$ and $\varepsilon_2$ vs $T$, obtained from the respective equivalent circuit fits. (b) Magnification of the intrinsic BiMnO$_3$ permittivity $\varepsilon_1$ displaying signs of MEC near $T_C$. (c) CPE exponents $n$, indicative of more ideal ($n \approx 1$) or less ideal ($n < 1$) character of the respective dielectric relaxation.

The extrinsic BiMnO$_3$ and BLNMO interface relative permittivity $\varepsilon_2$ displayed in Figure 24a shows values of 550 - 600 and 800 -1100 respectively, which is slightly below the standard for such type of relaxation [97]. Figure 24c shows the CPE exponent $n$ for the various BiMnO$_3$, BLNMO and BiFeO$_3$ relaxations. The intrinsic BiMnO$_3$ film relaxation shows a strongly increasing non-ideality above 75 K, which is manifested by a drop of $n$ with increasing $T$. This may well be consistent with a broadening of the distribution of dielectric relaxation times $\tau$, possibly as a reflection of the magnetic transition $T_C$. This, in turn, points towards increasing magnetic inhomogeneity in the BiMnO$_3$ film with increasing $T$ due to the nearby onset of paramagnetism above 75 K. The intrinsic BLNMO relaxation shows a similar decrease in the CPE exponent $n$ near $T_C$ as the BiMnO$_3$ film and the equivalent interpretation may apply.

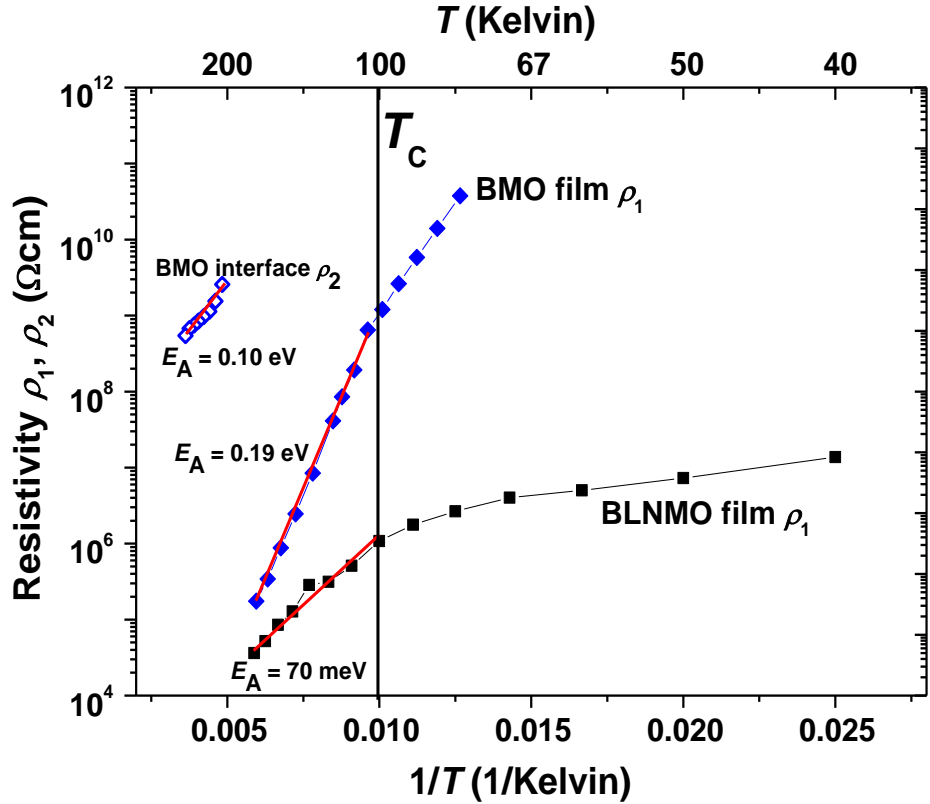

Figure 25. BiMnO$_3$ and BLNMO intrinsic and extrinsic resistivity $\rho_1$ and $\rho_2$ vs $T$, obtained from the respective equivalent circuit fits. Both the BiMnO$_3$ and BLNMO trends of $\rho_1$ vs $T$ show correlation with $T_C$. The activation energies $E_A$ are given in [eV/meV]. Solid (red) lines are linear Arrhenius fits.

Figure 25 demonstrates the BiMnO$_3$ and BLNMO thin film and interface resistivity, $\rho_1$ and $\rho_2$ respectively on plots of $\rho$ vs reciprocal $T$. Both the intrinsic film BiMnO$_3$ and BLNMO charge transport curves ($\rho_1$ vs 1/$T$) show correlations with $T_C$ with clearly modified $T$-dependences above and below $T_C$:

- Above $T_C$ electron hopping is indicated with a linear activation energy $E_A$ of 0.19 eV (BiMnO$_3$) and 70 meV (BLNMO). In case of BiMnO$_3$ this is in excellent agreement with previous work (0.2 eV) [101].

- Below $T_C$, the thermal activation of charge transport is weaker as reflected by a reduced slope of the $\rho_1$ vs 1/$T$ curves implying a lower activation energy $E_A$.

The charge transport in $BiMnO_3$ and BLNMO seems to be based on similar mechanisms, which is most likely by electron hopping between B-site cations. The intrinsic resistivity $\rho_1$ of the $BiFeO_3$ film was high at all $T$ investigated and could not be resolved. The film quality may therefore be regarded satisfactory with minimized leakage. In the equivalent circuit fits, the resistors R2 for the $BiMnO_3$/BLNMO interface and R1 for the $BiMnO_3$, BLNMO and $BiFeO_3$ film contributions were always set to infinity in several cases where the impedance data did not allow resolving the high resistance.

### 3.4.3. Magnetic Field Dependent Dielectric Characterizations

After rationalizing the presence of extrinsic and intrinsic dielectric relaxations in $BiMnO_3$ and BLNMO, and the single intrinsic relaxation in $BiFeO_3$ films, MR and MC in all types of film were determined. The $f$ dependent $BiMnO_3$, BLNMO and $BiFeO_3$ film impedance was measured at selected fixed $T$ repeatedly under various fixed applied $H$ and the data were fitted to the adequate equivalent circuit models established previously to extract the MR and MC from each circuit resistor and capacitor separately. The complex impedance notation, $Z^* = Z' + i\, Z''$, is presented in Figure 26 in terms of -$Z''$ vs $Z'$ curves for the $BiMnO_3$ film at 95 K close to $T_C$, where in such plots a semicircle is expected for each relaxation and the semicircle diameter corresponds to the resistivity of the respective relaxation.

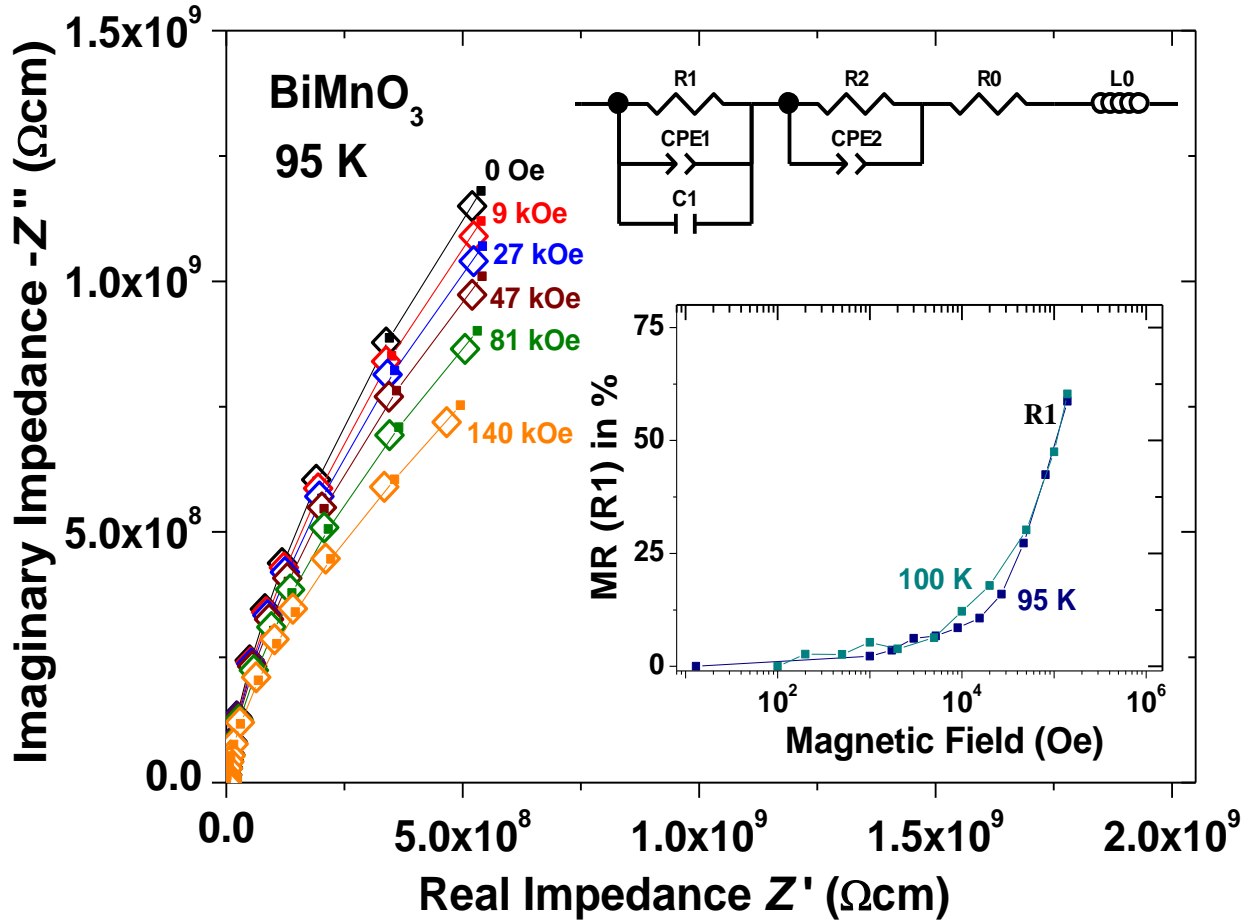

Figure 26. $BiMnO_3$ imaginary part of the impedance -$Z''$ vs real part $Z'$ at 95 K for selected applied magnetic fields $H$ as indicated. Open symbols (◊) represent experimental data, full squares (■) and solid lines represent fits to the data using the equivalent circuit depicted at the top of the curves. The decreasing size of the semicircle indicates MR of the intrinsic $BiMnO_3$ film relaxation. The inset displays MR vs $H$ at 95 and 100 K, where MR = [R($H$=0) - R($H$)]/R(0).

The partial semicircles in Figure 26 are a manifestation of the intrinsic relaxation, whereas the extrinsic interface resistance is large and not accessible at this $T$ and $f$ range. Although the intrinsic film semicircle is not fully displayed, the diameter may be extrapolated readily. On increasing $H$ the semicircle diameter decreases, which entails that $\rho_1$ decreases as a result of perceptible MR. The $H$ dependent R1 values obtained from the fittings allowed calculating the intrinsic MR in $BiMnO_3$ films defined as MR = [R($H$=0) - R($H$)]/R(0). The inset of Figure 26 shows rather high MR of up to 65% at 95 K and 100 K. No perceptible MR nor MC were detected above the $BiMnO_3$ magnetic transition $T_C$.

The BLNMO film showed similar MR as the $BiMnO_3$ film with a maximum in MR of ≈ 40% near the magnetic transition at ≈ 100 K [105]. In the $BiFeO_3$ films the resistance was too high to be resolved at all $T$ investigated as mentioned before, and the MR could not be determined.

Figure 27 shows $\varepsilon'$ vs $f$ curves for $BiMnO_3$ films collected under various applied $H$ at 95 K. At low $f$, $\varepsilon'$ shows an upturn, which represents the onset of the extrinsic Maxwell-Wagner relaxation R2-CPE2 (see $\varepsilon'_{high}$ in Figure 23a), where perceptible variations with $H$ occur. It has been pointed out before that such variation in $\varepsilon'$ with $H$ in the vicinity of an extrinsic contribution is not necessarily a reflection of MEC, but can be caused entirely by the MR of an extrinsic Maxwell-Wagner relaxation [95].

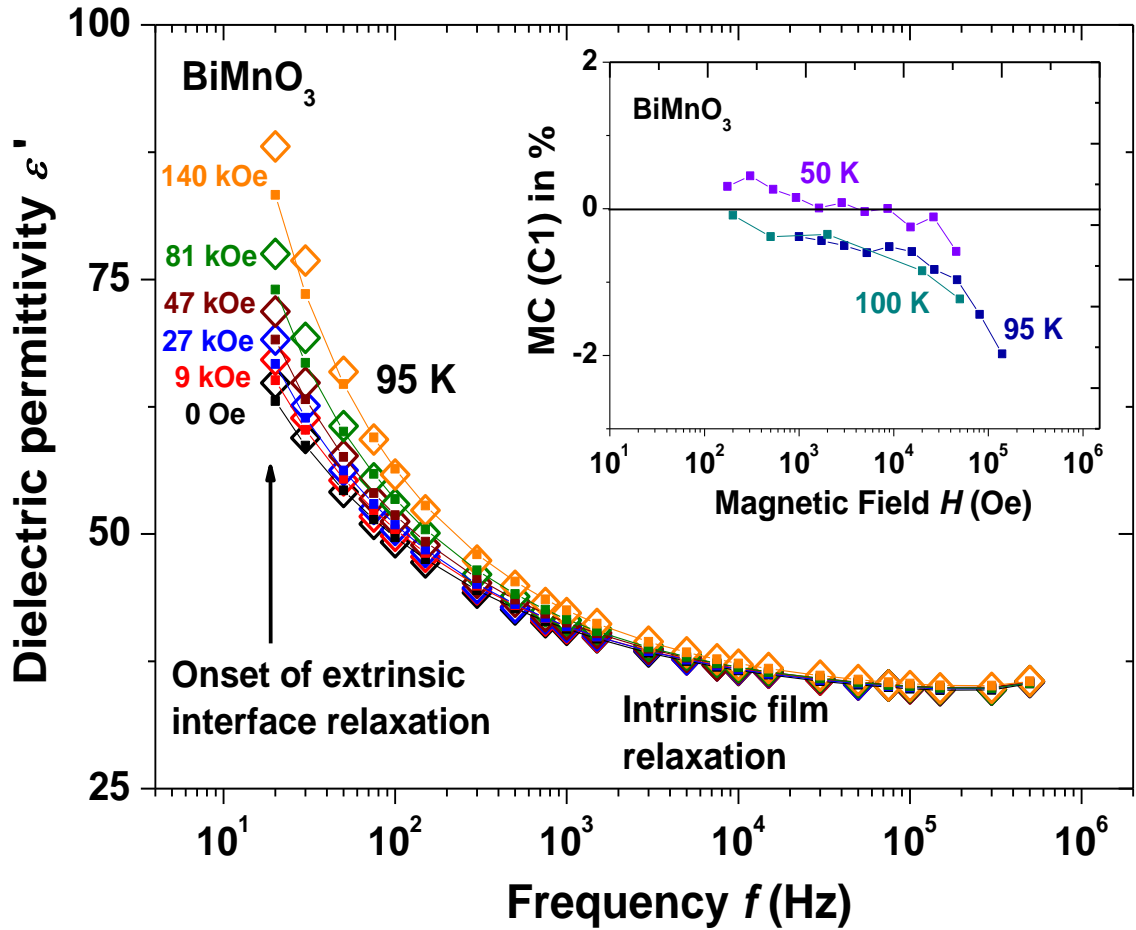

Figure 27. $\varepsilon'$ vs $f$ for a 50 nm $BiMnO_3$ thin film at 95 K for selected magnetic fields $H$ as indicated. Open symbols (◊) represent experimental data, full squares (■) and solid lines represent fits to the data using the equivalent circuit depicted in Figure 23a and 26. The inset displays MC vs $H$ at 50, 95 and 100 K, where MC = [C($H$=0) - C($H$)]/C(0).

Since the intrinsic BiMnO$_3$ film relaxation shows perceptible MR, the extrinsic relaxation may be expected to exhibit similar behavior and the variation in $\varepsilon'$ at low $f$ observed in Figure 27 may well be artificial and simply reflect the MR of the extrinsic Maxwell-Wagner relaxation. At higher $f$ where the intrinsic relaxation is more dominant, only small MC can be seen.

Intrinsic BiMnO$_3$ $\varepsilon_1$ values were obtained from the equivalent circuit fits by extracting C1 for various $H$ and the intrinsic MC, defined as MC = [C($H$=0) - C($H$)]/C(0), is shown in the inset of Figure 27.

Intrinsic MC and MEC appear to be rather small in the range of $\approx$ -1.5% at 90 kOe, which confirms the rather weak peak feature in $\varepsilon_1$ vs $T$ (Figure 24b) and previous reports on polycrystals (MC $\approx$ -0.7% at 90 kOe) [22]. In the BLNMO film only signs of an extrinsic and artificial MEC due to MR effects were detected, whereas BiFeO$_3$ films exhibited no MEC at all. In BiFeO$_3$ the absence of any MEC may be debited to the fact that the antiferromagnetic transition temperature occurs at much higher $T$ than the range investigated here, and only near such transition MC and MEC effects may be detectable.

## CONCLUSION

It was shown that deposition of multiferroic epitaxial thin films based on the Bi containing perovskite oxides BiMnO$_3$, (Bi$_{0.9}$La$_{0.1}$)$_2$NiMnO$_6$, and BiFeO$_3$ can be performed adequately by pulsed laser deposition. Structural film characterizations demonstrated the desired single film phases, single out-of-plane orientation and coherent in-plane strain. BiFeO$_3$ films were shown to be ferroelectric using PFM at room temperature and strong indications for ferroelectricity were observed in BiMnO$_3$. All films showed the desired ferromagnetic (BiMnO$_3$) or ferrimagnetic (BiFeO$_3$ and (Bi$_{0.9}$La$_{0.1}$)$_2$NiMnO$_6$)) structure. Furthermore, it was shown that magnetic field dependent impedance spectroscopy is a powerful tool to deconvolute intrinsic and extrinsic contributions to the film dielectric properties, which concomitantly allows reliable determination of intrinsic magneto-electric coupling effects. Such effects were shown to be absent in BiFeO$_3$ and double-perovskite (Bi$_{0.9}$La$_{0.1}$)$_2$NiMnO$_6$ at the temperature range investigated, whereas BiMnO$_3$ films showed weak MEC effects of up to 2% near the magnetic transition $T_C \approx$ 100 K.

## ACKNOWLEDGMENTS

E.L., J.V. and M.V. acknowledge financial support from the Spanish Ministerio de Ciencia e Innovación (MICINN) under project numbers MAT2008-06761-C03, MAT2011-29269-C03 MAT2011-27470-C02-01, MAT2011-27470-C02-02, IMAGINE CSD2009-00013 and NANO-SELECT CSD2007-00041. RS acknowledges the Ramón y Cajal program from the MICINN. Many thanks go to SSSS Ltd. for support with the data analysis. Furthermore, the contributions from Carmen Munuera, Neven Biskup, Norbert Nemes, Alberto Rivera, Mar García-Hernandez, Carlos León, Josep Fontcuberta and Jacobo Santamaría are acknowledged explicitly.

Bi Containing Multiferroic Perovskite Oxide Thin Films    37